\documentclass[openacc]{rstransa}

\usepackage{marvosym} 
\usepackage{wasysym}
\usepackage{color}

\renewcommand{\H}{\operatorname{H}}
\newcommand{\I}{\operatorname{I}}

\newcommand{\eM}     {$\epsilon$-machine}
\newcommand{\eMs}    {$\epsilon$-machines}

\newcommand{\EMs}    {$\epsilon$-Machines}

\newcommand{\hmu}		{ {h_\mu}}
\newcommand{\EE}		{ {\bf E}}

\newcommand{\rhomu}		{ {\rho_{\mu}}}
\newcommand{\bmu}		{ {b_{\mu}}}
\newcommand{\rmu}		{ {r_{\mu}}}
\newcommand{\qmu}		{ {q_{\mu}}}

\newcommand{\eg} {{\it {e.g.}}}
\newcommand{\ie} {{\it {i.e.}}}
\newcommand{\etal} {{\it {et~al.}}}

\newcommand{\kB} { k_\text{B} }

\newcommand{\iceh}{ice I$_{\rm{h}}$}
\newcommand{\icec}{ice I$_{\rm{c}}$}
\newcommand{\icesd}{ice I$_{\rm{sd}}$}
\newcommand{\icech}{ice I$_{\rm{ch}}$}

\begin{document}

\title{What did Erwin Mean?\\
The Physics of Information from the Materials Genomics of
Aperiodic Crystals \& Water to Molecular Information Catalysts \& Life}

\author{D.~P.~Varn and J.~P.~Crutchfield}

\address{Complexity Sciences Center \& Physics Department,
University of California,
One Shields Ave,
Davis, CA 95616}

\subject{Aperiodic crystal, information theory, computational
mechanics, nonequilibrium thermodynamics, crystallography}


\corres{D.~P.~Varn, J.~P.~Crutchfield\\
\email{dpv@complexmatter.org, chaos@ucdavis.edu}}

\begin{abstract}
Erwin Schr\"{o}dinger famously and presciently ascribed the vehicle
transmitting the hereditary information underlying life to an `aperiodic
crystal'. We compare and contrast this, only later discovered to be stored in
the linear biomolecule DNA, with the information bearing, layered
quasi-one-dimensional materials investigated by the emerging field of
\emph{chaotic crystallography}. Despite differences in functionality, the same
information measures capture structure and novelty in both, suggesting an
intimate coherence between the information character of biotic and abiotic
matter---a broadly applicable physics of information.  We review layered solids
and consider three examples of how information- and computation-theoretic
techniques are being applied to understand their structure.  In particular, (i)
we review recent efforts to apply new kinds of information measures to quantify
disordered crystals; (ii) we discuss the structure of ice I in
information-theoretic terms; and (iii) we recount recent experimental results
on tris(bicyclo[2.1.1]hexeno)benzene TBHB), showing how an
information-theoretic analysis yields additional insight into its structure. We
then illustrate a new Second Law of Thermodynamics that describes information
processing in active low-dimensional materials, reviewing Maxwell's Demon and a
new class of molecular devices that act as information catalysts. Lastly, we
conclude by speculating on how these ideas from informational materials science
may impact biology.
\end{abstract}

\begin{fmtext}
\end{fmtext}

\maketitle

\section{Introduction}
\label{Sec:Introduction} 

To account for the `special properties' of life---\eg, movement, metabolism,
reproduction, development---the prevailing wisdom from the time of Aristotle
into the 19th century was that organic matter differed in some fundamental way
from inorganic matter. While this notion, called \emph{vitalism}, may seem
quaint to 21st century scientists, it held sway until the chemist Friedrich
W\"{o}hler showed that, unexpectedly, a known organic compound, urea, could be
artificially synthesized from cyanic acid and ammonia~\cite{Kinn99a}. This
fabrication process, while different than that used in biological systems,
nonetheless served as an important clue that the divide between living and
nonliving matter was not absolute. Abiotic processes could make substances
theretofore only encountered in biologically derived materials.  Additionally,
we see that---and not for the last time---results obtained from one discipline,
chemistry, have had important consequences in another, biology. This confluence
of diverse avenues of inquiry coalescing into an ever larger conceptual
picture of Nature is, of course, an oft-repeated theme in the sciences. Other
famous examples include Newton's discovery that the motion of celestial bodies,
such as the moon planets, and that of terrestrial ones under the influence of
gravity, such as the proverbial apple, both are manifestations of a universal
law of gravitational attraction; James Clerk Maxwell's unification of
electricity and magnetism into his famous equations; and James Prescott Joule's
demonstration that the caloric was nothing but energy by another name, now
formalized in the First Law of Thermodynamics.  Indeed, E.~O.~Wilson takes the
extreme position that \emph{all} human knowledge, from the most concrete of the
sciences to the least precise of the liberal arts, is ultimately
interlinked~\cite{Wils98a}.

We need not go quite so far as Wilson. It is enough for our purposes to realize
that while `abiotic' sciences such as physics, chemistry, astronomy, and
geology share obvious strong interconnections, biology has remained relatively
aloof.  This is not to say that biology has not benefited greatly from
knowledge transferred to it from other physical sciences. In addition to the
urea example above, note that metabolism is at its core a question of the
utilization and transformation of energy---a notion made concrete and
operational in physics. And too, biology has benefited tremendously from
techniques and discoveries made in other sciences. Indeed, in 1937 Max
Delbr\"uck (Nobel Prize Physiology or Medicine 1969) adapted his training in
astrophysics and theoretical physics to probe gene susceptibility to mutations,
stimulating physicists' interest in biology and establishing molecular biology.
More familiar, though, it was the infamous X-ray diffraction image known as
`photograph 51' from the lab of Rosalind Franklin that provided a key insight
leading geneticist James Watson and physicist Francis Crick (Nobel Prize in
Physiology or Medicine 1962) to propose the double helical structure of
DNA~\cite{Wats53a}. Despite the above, biology is clearly the least well
integrated in the family of sciences. We can speculate that the sheer
complexity of life and the novel phenomena it displays are at least partially
responsible for this. Even one of the most basic organisms, \emph {Mycoplasma
genitalium}, has genome of `only' $580,070$ base pairs~\cite{Fras95a}. Biology
is complicated.

And it is perhaps due to this complication that the mathematical `sciences'\footnote{We
                 take the view that science is fundamentally an experimental
		endeavor, ultimately dependent on empirical observation. Pure
		mathematics, while of enormous interest for both its intellectual vigor
		and beauty, as well as its practical applications, need not make appeal
		to experiment for validation or refutation of its claims and, thus,
		does not constitute a scientific discipline.}
have made their least impact in theoretical biology. By and large, the advanced
mathematical techniques that saturate any theoretical physics text find no
counterpart in biology texts. There is one area, however, where arguably
biology has outpaced her sister sciences: the incorporation of
\emph{information theory}\cite{Shan48a,Cove06a} into the description of
physical systems. And, we will suggest that biology has carved a conceptual
path that abiotic physical sciences would do well to emulate. Before we move
too far ahead, though, let's start at the turn of the 20th century and visit
one of the many revolutionary advances that ushered in the era of `modern'
physics and that remains today a key probe of molecular biological structure. 

\section{Structure, Aperiodic Crystals, and Information}

The immense conceptual advances in physics made in the first third of the 20th
century are legion, but here we focus on the contributions to the structure of
matter. While it is Max von Laue (Nobel Prize Physics 1914) that is credited
with the discovery of the diffraction of X-rays by crystals, it is the father
and son team, Sir William Henry Bragg and William Lawrence Bragg (Nobel Prize
Physics 1915), that receive much of the credit for exploiting it as a tool to
determine crystal structure. For a periodic repetition of some pattern, as one
might find in the simple crystals such as NaCl, the diffraction pattern is
dominated by very strong reflections at particular angles, called \emph{Bragg
reflections}. Much weaker diffuse scattering is known to occur between the
Bragg reflections and had been observed as early as 1912 by Walter Friedrich.
While this diffuse scattering can be explained by the thermal motion of the
constituent atoms, it could genuinely be a harbinger of deviations from perfect
periodic order. But the assumption of periodicity greatly simplifies the
analysis of diffraction patterns, and the early years of crystallography were
marked with enormous success in solving for the periodic structures that seemed
so common. Indeed, it may be argued that this research program so successful in
describing a particular kind of structure---periodic structure, ``the infinite
repetition in space of identical structural units''---came at the cost of
developing alternate theoretical tools.

On the biology front, cognizant of Delbr\"uck's results on mutations, the
prominent physicist Erwin Schr\"{o}dinger (Nobel Prize Physics 1933) was busy
considering life from a physics point-of-view. In his now classic 1944 book,
\emph{What is Life?}~\cite{Schr44a}, Schr\"{o}dinger introduces two concepts
that are of interest to us here. The first is \emph{negentropy}, or the entropy
that an organism exports to its surroundings to keep its internal entropy low.
If one views entropy as a measure of disorder, then the Second Law of
Thermodynamics makes it clear that for an organism to maintain some structure,
it must rid itself of the disorder that accompanies life-maintaining processes.
The second, and equally important, is the idea that the hereditary mechanism
that must exist so that traits of individuals can be passed to offspring could
be housed in what he called an \emph{aperiodic crystal}. Although H.~L.~Muller
made a similar proposal over twenty years previous, it was Schr\"{o}dinger's
advocacy that captured the imagination of Crick and Watson to seriously
investigate this possibility. Schr\"{o}dinger's aperiodic crystal was some
material substrate, perhaps a molecule, that lacked strict periodicity. The
reason for this is that exact repetition of a motif, in other words a crystal,
is information poor---too poor to carry heredity. Without some unpredictability,
or novelty, nothing new is learned and communicated. It is remarkable that
Schr\"{o}dinger made this prediction before a quantitative understanding of
information was articulated.

In 1947, three physicists from Bell Telephone Laboratories, John Bardeen,
Walter Brattain, and William Shockley (Nobel Prize Physics 1956) invented a
small device that revolutionized the design of electrical circuits: the
transistor, which ushered in the era of electronics. It's significance was
immediately recognized and a press release was duly issued the next year.  Yet,
arguably~\cite{Glei11a}, this was only the \emph{second} most important
announcement to come out of Bell Labs in 1948. The first came from a thirty-two
year-old mathematician, engineer, and cryptographer, Claude E.~Shannon\footnote{In a perhaps unexpected overlap to
	our narrative, Shannon's PhD thesis from Massachusetts Institute of
	Technology (1940) was titled \emph{An Algebra for Theoretical Genetics}
	and explored the mathematics of genetic recombination.},
in the form of a paper in the \emph{Bell System Technical Journal} with the unassuming title ``A Mathematical Theory of Communication"~\cite{Shan48a}.

Shannon's main premise is that information is a degree of surprise. Given an
\emph{information source} $X$---a set of messages $\{x\}$ that occur with
probabilities $\{\Pr(x)\}$---an individual message's \emph{self-information} is
$\H(x) = - \log_2 \Pr(x)$. Thus, predictable events ($\Pr(x) = 1$) are not
informative---$\H(x) = 0$, since they are not surprising. Wholly unpredictable
events, such as the flips of a fair coin, are highly informative:
$\H(\text{Heads}) = - \log_2 \tfrac{1}{2} = 1$. When using logarithms base $2$
the information unit is a \emph{bit} or binary digit. Shannon's first major
result was to show that the average self-information, what he called the
\emph{entropy} paralleling Boltzmann and Gibbs in vocabulary and notation,
$\H[X] = - \sum_{x \in X} \Pr(x) \H(x)$ measures how compressible a source's
messages are. However, quantifying information was simply preliminary to
Shannon's main motivation. Working for the Bell Telephone Company, a
communications enterprise, his main goal was to lay out operational constraints
for communicating information over noisy, error-prone transmission equipment,
which he formalized as a \emph{communication channel}. The result was his most
famous and far-reaching result: As long as the source entropy is less that the
channel's transmission capacity---$\H[X] < \mathcal{C}$---then, even if errors
are introduced there is a way to encode the source messages such that the
receiver observing the noisy channel output can \emph{exactly} reconstruct the
original messages.  This single result is key to almost all communication
technologies that drive today's modern economies.

Shannon himself was rather careful to distance his quantitative theory on the
amount of information in a source from discussions of that information's
meaning or semantic content \cite{Shan56b}. His goal was the operational result
just recounted that did not require knowing what information was being
communicated.  However, as we will explain, his measure of information and its
semantics turn out to provide a central and quantitative tool for understanding
the organization of materials that are more than periodic crystals---materials
that are not regular repetitions of identical unit cells. We call this
application of Shannon's information theory to material structure ``chaotic
crystallography'', for reasons that will become evident.

What kinds of materials are not crystals? An obvious class is those in which
atoms of random kinds are randomly placed in space. The resulting
dichotomy---materials are either periodic or random---is too simple a view.
There is a spectrum. A first example, one controversial in its time, came in
the discovery of quasi-crystals~\cite{Shec84a}: Metals with long-range
orientational order, an icosahedral phase, but no translational symmetry. This
fell so far outside of the periodic-random dichotomy that it was some years
after experimental detection that quasi-crystals were widely accepted (Nobel
Prize in Chemistry 2011).

We now know that the spectrum of material structures between periodic crystals
and amorphous materials is populated by much more than quasi-crystals. First
are the so-called \emph{aperiodic crystals} that exhibit sharp diffraction
peaks, but not lattice periodicity ~\cite{URL_DefintionOfAPeriodicCrystal}.
The mathematics of such aperiodic orders has blossomed \cite{Baak13a}.  Many
disordered materials, however, exhibit broadband diffraction patterns but also
have a large degree of organization. These are what we now call \emph{chaotic
crystals}. The aperiodic crystals, by way of comparison, are seen to lie in the
narrow boundary region between periodic and chaotic crystal materials.

Given this wide spectrum, one needs tools that readily describe processes that
range from periodicity to randomness and capture the intermediate semi-ordered,
semi-disordered structures. Information theory is one of those tools. We will
describe how it applies to material structure, forming the endeavor of chaotic
crystallography. A compelling insight is that though we start with a focus just
on surprise and prediction, we are led to novel notions of structure, partial
symmetries, and information storage.

\section{From Information Measures to Structure}
\label{InformationMeasures}

Although fundamental to the practice of science, a thorough understanding of
the information obtained from individual measurements has only recently been
examined in detail~\cite{Jame11a,Jame14a}. The key issues at hand are easily
stated: Given a history of such measurements, how much does one learn from any
particular observation?  How much of the past is useful for predicting the
results of future measurements? To what degree is a measurement simply
randomness, and not structure? How much information obtained in the present is
transmitted to the future?  Perhaps not surprisingly, considering these
questions in the light of information theory~\cite{Cove06a} revealed a number
of new computational and informational measures that give important insights
into how correlations are manifested in different kinds of structure.

\begin{figure}
\begin{center}
\includegraphics{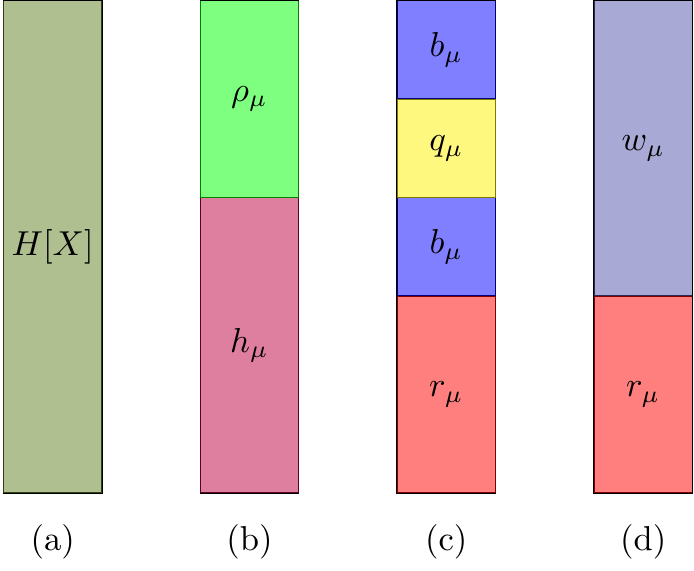}
\caption{{\small {Dissecting the information $\H[X]$ learned from a
  measurement of a single random variable $X$.
  (Figure from James \etal\ \cite{Jame11a}, used with permission.) }}}
\label{Fig:Dissection}
\end{center}
\end{figure}

As noted above, the workhorse of information theory is the \emph{Shannon
entropy}~\cite{Shan48a} of a random variable $X$: $\H(X) = -\sum_{x \in X}
\Pr(x)\log_2\Pr(x)$, where the $x$ are the possible realizations of the
discrete variable $X$ and $\Pr(x)$ is the probability of observing $x$.  While
Shannon entropy has many interpretations, most useful here is that it is the
average amount of information an observation reveals when measuring that
variable. Real measurements are often sequential in time and one might expect
that there are correlations between measurements. The extension of the Shannon
entropy to a series of measurements follows naturally by replacing the single
random variable $X$ with the sequence $X_1, X_2, \dots X_{L}$ of random
variables---often written $X^{L}$---and the realization of a single measurement
$x$ by the series of measurements $x_1, x_2, \dots x_L$, the latter
conveniently denoted $x^{L}$. Thus, by considering successively longer pasts
---$X_{-1}$, then $X_{-2} X_{-1}$, and so on---one can quantify how less
uncertain a measurement of $X_0$ is. Or, stated differently, we can quantify
how much knowledge of the past reduces the information learned in the present:
$\H[X_0] \geq \H[X_0|X_{-1}] \geq \H[X_0|X_{-1},X_{-2}] \geq \ldots$, where we
introduced the conditional Shannon entropy $\H[X|\cdot]$. It is also useful to
consider the entropy rate $\hmu$, the information learned on average per
observation, having seen an infinite past: $\hmu = \H[X_0 | \ldots X_{-3}
X_{-2} X_{-1}]$.

Since information theory was originally developed in the context of
communication, imagined as a temporal progression of symbols, a natural notion
of past, present, and future permeated the theory.  Operating under this
prejudice introduced a preferred arrow of time.  As a consequence, the utility
of conditioning \emph{current} measurements, or observed symbols, on
\emph{future} observations was not obvious. From a mathematical point of view,
of course, there is no inherent impediment to doing this. However, replacing a
time series by a spatial one lifts the directional prejudice, opening a way to
identify other measures of information that treat the past and future on equal
footing~\cite{Jame11a,Elli09a,Crut09a,Maho09a,Maho11a}.

As an example, consider a single measurement of the random variable $X$. The
theoretical maximum amount of information that one can possibly learn is just
$\H[X]$, see Fig. (\ref{Fig:Dissection}a).
However, if there are correlations or regularities in the data, some
of this could have been anticipated from previous observations. Let us call
this part the \emph{redundancy rate} $\rhomu = \I[X_0:\ldots X_{-3} X_{-2}
X_{-1}]$---the \emph{mutual information} between the present $X_0$ and the past
$\ldots X_{-3} X_{-2} X_{-1}$. The other part of the information could not be
anticipated; it truly is random and is just $\hmu$. Thus, the amount of
information available in a measurement naturally decomposes into these two
parts, as shown in Fig. (\ref{Fig:Dissection}b).

However, further conditioning yields further decomposition of each of these.
First, the random portion $\hmu$ breaks into two parts: the \emph{ephemeral
information rate} $\rmu$ and the \emph{bound information rate} $\bmu$. The
ephemeral information rate $\rmu = \H[X_0|\ldots X_{-3} X_{-2} X_{-1}, X_{1}
X_{2} X_{3} \ldots ]$ is the information that exists only in the present. It is
not predictable from the past nor is it communicated to the future. Existing
only in the present, it is ephemeral. The bound information rate $\bmu =
\I[X_0 : X_{1} X_{2} X_{3} \ldots | \ldots X_{-3} X_{-2} X_{-1}]$ is the
information shared between the present and future, but is not in the past. As
such, it measures the rate at which spontaneously generated information
($\hmu$) is actively stored by a system. Second, the redundancy rate also
breaks into two parts, the first again being $\bmu$ and a second part called
the \emph{enigmatic information rate} $\qmu$. The latter is three-way mutual
information $\I[\ldots X_{-3} X_{-2} X_{-1}: X_0: X_{1} X_{2} X_{3} \ldots]$
shared between the past, the present, and the future.

The net ``decomposition'' of the information $\H[X_0]$ in a single measurement
is illustrated in Fig.~(\ref{Fig:Dissection}c). This is only a sampling of the
possible ways that information can be semantically partitioned between the
past, present, and future. Figure~(\ref{Fig:Dissection}d), for example, is a
decomposition into dissipated $\rmu$ and useful information $w_\mu$. Moreover,
other additional measures, discussed by James
\etal~\cite{Jame11a,Jame14a,Ara15a}, have been defined and explored.
Importantly, they now can all be analytically calculated from a process's \eM\
\cite{Crut13a,Riec14a}, once that is in hand.

\section{Chaotic Crystallography}
\label{sec:ChC}

Armed with this new arsenal of structural information measures, a detailed,
quantitative picture of how information is shared between the past, present,
and future is made plain. With these in mind, \emph{intrinsic computation} is
defined as how systems store, organize, and transform historical and spatial
information \cite{Crut89a,Feld08a}. Different processes may have quantitatively
and qualitatively different kinds of intrinsic computation, and understanding
these differences gives insight into how a system is structured \cite{Crut94a}.

\emph{Chaotic crystallography} (ChC)
\cite{Riec14a,Varn13a,Varn02a,Varn04a,Varn07a,Varn13b,Riec14c} then is the
application of these information- and computation-theoretic methods to discover
and characterize structure in materials. It reinterprets the time axis, used
above for pedagogical reasons, for a one-dimension spatial coordinate along
some direction in a material. The choice of the name is intended to be
evocative: we retain the term ``crystallography'' to emphasize continuity with
past goals of understanding material structure; and we introduce the term
``chaotic'' to associate this new approach with notions of disorder,
complexity, and information processing. Using chaotic crystallography we can describe the ways in
which this information decomposition quantitatively captures crystal
structure---distinguishing structure that might be expected, \ie, repetitive
periodic structure, from that structure not expected, \ie, faulting structure.

\subsection{Material Informatics of Faults and Defects}

Since classical crystallography~\cite{Dorn56a,Mack86a,Mack95a} largely concentrates on periodic structures, it encounters difficulty classifying structures that do not fit this paradigm. Most efforts have centered on describing how a crystal, that presumably could have been perfectly ordered, falls short of this ideal. For example, in close-packed structures, Frank~\cite{Fran51a} distinguished two kinds of layer faults: {\emph {intrinsic}} and {\emph {extrinsic}}. For intrinsic faults, each layer in the material may be thought of as belonging to one of two crystal structures: either that to the left of the fault or that to the right. It is as if two perfect, undefected crystals are glued together and the interface between them is the fault. In contrast, it may be that a particular layer cannot be thought of as a natural extension of the crystal structure on either side of the fault. These are extrinsic faults. Another classification scheme has its origins in the mechanism that produced the fault. In close-packed structures, commonly encountered faults include {\emph {growth faults}}---\ie, those that occur during the crystal growth process; {\emph {deformation faults}}---which are often associated with some post-formation mechanical stress to the crystal; and {\emph {layer-displacement faults}}---which can occur by diffusion between adjacent layers. As each is defined in relation to its parent crystal structure, each kind of crystal structure typically has its own distinctive morphology for each kind of fault.

The result is a confusing menagerie of stacking sequences that deviate from
the normal. This collection may not be exhaustive, depending on how large of a
neighborhood one considers nor may particular sequences be unambiguously
assigned to a particular kind of fault structure. Indeed, in the event that
there are multiple kinds of faults, or multiple mechanisms for producing
faults, an attempted analysis of the fault structure may be indeterminate~\cite{Varn02a}.
Faulting may also be classified in terms of how faults are spatially related to
each other. The absence of correlation between faults implies \emph {random
faulting}. Alternatively, the presence of a fault can influence the
probability of finding another fault nearby. This latter phenomenon is called
\emph {nonrandom} faulting and is not uncommon in heavily defected specimens.
Lastly, in some materials faults appear to be regularly interjected into the
specimen, and this is referred to as \emph {periodic faulting}. Screw
dislocations are thought to be a common cause for these latter
faults~\cite{Seba94a}.

\begin{figure}
\begin{center}
\includegraphics{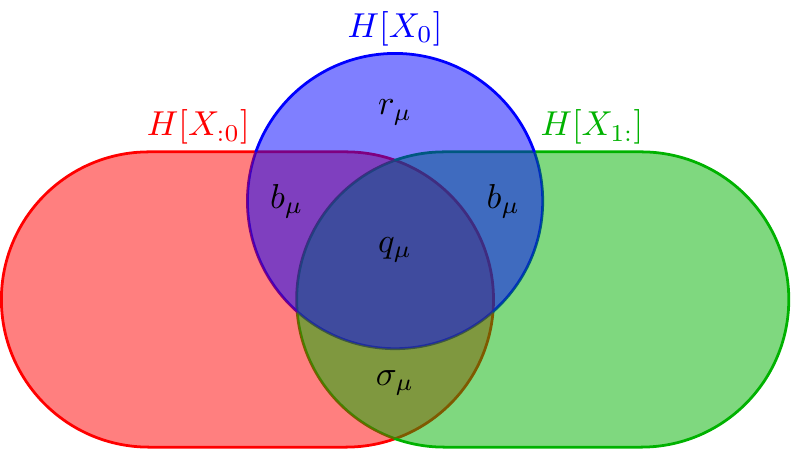}
\caption{{\small {Information diagram showing the information anatomy of
  $\H[X_0]$ in the context of the full spatial stacking of layers in a chaotic
  crystal. Let $X_0$ be the layer of interest, $X_{:0}$ be an arbitrarily long
  but finite section of the specimen to the left of $X_0$, and $X_{1:}$
  similarly be an arbitrarily long but finite section of the specimen to the
  right of $X_0$.  The left $X_{:0}$ partitions $\H[X_0]$ into two pieces:
  $h_{\mu}$ and $\rho_{\mu}$. The right $X_{1:}$ then partitions those into
  $r_{\mu}$, two $b_{\mu}$s and $q_{\mu}$. (Recall Fig. \ref{Fig:Dissection}
  which decomposed only $\H[X_0]$.) This leaves a component $\sigma_{\mu}$, the
  \emph{elusive information}, that is shared by the left and right, but is not
  in the present layer. When positive, it indicates that not all of the
  correlation between left and right half-configurations is contained in
  locally and so there are internal hidden mechanisms that carry the
  correlation \cite{Ara15a}.
  (Figure from James \etal\ \cite{Jame11a}. Used with permission.) }}}
\label{Fig:Anatomy}
\end{center}
\end{figure}

These phenomenological categorizations, while often helpful and sensible,
especially for weakly faulted crystals, are not without difficulties. First, it
is clear that each is grounded in the assumption that the native, or ideal,
state of the specimen must be a periodic structure. This bias, perhaps not
intentionally, relegates nonperiodic stacking to less stature, as is evident in
the use of the term ``fault''. It may be rather that disorder is the natural
state of the specimen~\cite{Remp10a}, in which case employing a framework that
incorporates this feature of matter upfront will prove more satisfactory. In
fact, it is not even clear that periodic order should be the ground state for
many kinds materials, even for those with finite-range interactions and in the
absence of fine-tuning of energetic coupling parameters between
layers~\cite{Varn01a}, as is found in axial next-nearest neighbor Ising (ANNNI)
models~\cite{Yeom88a}. Second, an analysis of the stacking structure based on
these categories may not be unambiguous, especially in the case of heavy
faulting. Third, this entire view is only tenable in the limit that a parent
crystal exists; \ie, it only applies in the weak faulting limit.

Consistency can be brought to this complicated picture of material structure by
using information theory \cite{Varn15a}. A complementary view may be postulated
by asking how information is shared and distributed in a crystal, and a natural
candidate for this kind of analysis is to employ the information measures
above.  Although the previous exposition used a temporal vocabulary of a past,
present, and future, there is no mathematical change to the theory if instead
we adopt the view that the observed sequences are spatial configurations. That
is, there are measurements that are to the left of the present measurement, the
present measurement itself, and those measurements to the right of the current
measurement.  For quasi-one-dimensional materials we think of each
measurement as the orientation of a layer. This view of a sequence of layer
orientations translates to an \emph{information diagram} or \emph{$I$-diagram},
as shown in Figure~\ref{Fig:Anatomy}. There, we see how information is shared
between the different halves of the specimen and the current layer. The
information measures given in terms of mutual information can be interpreted as
\emph{layer correlations} within in the specimen. Importantly, although one
typically averages them over the crystal, it is possible instead to not perform
that average, but examine them layer-by-layer. As shown in James \etal\
\cite{Jame14a}, information-theoretic measures can be quite sensitive to changes in system parameters and
we expect will provide a barometer quantifying important aspects of material structure.

As an example, electronic structure calculations arising from one-dimensional
potentials are known to depend on pairwise correlations~\cite{Ande58a,Izra12a}, with
the transmission probability spectrum of an electron through such potentials
often governed by the correlation length.  Information-theoretic quantities,
with their more nuanced view of correlation lengths in terms of conditional and
mutual informations, give a more detail picture of the role of disorder in
electronic structure. One of the simpler and more common measures of all-way
correlation is the mutual information between the two halves of a specimen: the
\emph{excess entropy} $\EE = \I[\ldots X_{-3} X_{-2} X_{-1}: X_0 X_1 X_2
\ldots]$. Inspection of the information diagram reveals its decomposition into
information atoms: $\EE = \bmu + \qmu + \sigma_\mu$.

Additionally, not only is the global structure important, but local defects can
introduce local deviations from average structure, as seen in Anderson
localization~\cite{Izra12a}. This is a current area of research interest~\cite{Lei15a}.
Similarly, regions of charge surplus or depletion can effect other properties,
such as the transmission of light. The area of disordered photonics attempts to
understand and exploit such structures for new technologies~\cite{Wier13a}.

Thus, a number of questions can be asked concerning the distribution of
information in the crystal as revealed in its structure.  For example, how much
information is obtained from the current measurement? Is this shared with its
neighbors or is it localized? Considering questions such as these leads to a
new categorization of disordered structure in crystals.


\subsection{Chaotic Crystals: Structure in Disorder}

The net result is a consistent, quantitative, and predictive theory of
structure in disordered materials that extends beyond faulting and weak
disorder and that applies to the full spectrum of material structure from ideal
periodic crystal to amorphous materials and complex long-ranged mixtures in
between. As Ref. \cite{Ball14a} notes, in short, we have a new view of what
crystals are and can be. Reference \cite{Varn15a} reviews how this works in
detail.

The term ``chaotic crystal'' has been used in two previous contexts. In 1991
Leuschner~\cite{Leus91a} introduced several models of structure for
one-dimensional crystals, capable of producing completely periodic,
quasiperiodic, and chaotic behavior. The latter was accomplished using the
Logistic Map~\cite{Crut89a} as a generator of uncertainty in the stacking
sequence---in effect using it as a random number generator. Later, Le Berre
\etal~\cite{Lebe02a}, in the context of steady-state pattern formation of
two-dimensional systems, defined a chaotic crystal as ``any structure without
long range order, but spatially statistically homogeneous''. Our use of the
term is both less restrictive, in accounting for long-range order, and more
general, in allowing for a wide range of types of disorder. It should be
apparent that the chaotic crystal we describe here is just the kind of crystal
that Schr\"{o}dinger imagined as the carrier of heredity. While he called it an
aperiodic crystal, that term has been usurped to describe a very special kind
of deviation from periodicity, the kind that is found to preserve sharp peaks in
the diffraction pattern~\cite{URL_DefintionOfAPeriodicCrystal}. Thus, we use
the term chaotic crystal to indicate a broader notion of noncrystallinity, one
that encompasses structures with a nonzero entropy density, as is needed for
any structure to house information.

Let's illustrate how chaotic crystallography applies to real-world
materials---the closed-packed structures of ice and a complex molecule used to
probe the chemistry of benzene's aromaticity. Then, combining these results
with previous chaotic crystallographic analyses of zinc sulfide (ZnS), we demonstrate
how a unified vision of organization in materials is emerging.

\subsubsection{Layer Disorder in Ice I}

Although often thought of as merely the medium of life---albeit an essential
one\footnote{The necessity of water to life has come under significant
scrutiny~\cite{Ball08a}.}---there has been growing appreciation of the active
role that water plays in biological processes. As an example,
Ball~\cite{Ball08a,Ball08b} cites the generic interaction of two proteins.  If
both are dissolved in the cellular medium, the intervening water molecules must
be removed for an interaction to occur. Water is, of course, polar, and
displacing the last few layers of water may be nontrivial, depending on for
instance to what degree the protein activation sites are either hydrophilic or
hydrophobic. Additionally, one should expect properties of thin water films,
such as viscosity, to deviate significantly from their bulk properties. Even
the simulation of complex polypeptides is incomplete without considering the
influence of water~\cite{Ball08a}. As another example, there is evidence that
life engineers and precipitates the formation of ice. Without the influence of
impurities to act as centers of inhomogeneous ice nucleation, water in clouds
can be expected to freeze at 235~K via homogeneous ice nucleation. Impurities
such as soot, metallic particles, and biological agents can raise this
temperature. A particularly effective biological agent is the bacterium
\emph{Pseudomonas syringae} that, due to protein complexes on its cell surface,
can initiate freezing at temperatures as high as 271~K~\cite{Atta12a}.
Although its particular role may be highly varied depending on circumstances,
regarding water as merely ``the backdrop on which life's molecular components
are arrayed"~\cite{Ball08a} is quite untenable.

Given the structural simplicity of a water molecule---H$_2$O---and its
importance to biological as well as other natural systems, it is perhaps
surprising that, in both its liquid and solid forms, H$_2$O remains somewhat
mysterious. In the liquid state, water molecules form ``networks", where the
connections are made from hydrogen bonds, giving the substance considerable
structure. So too, ice shows considerable and variable structure. There are no
less than fifteen known distinct polymorphs of ice (usually specified by Roman
numerals)~\cite{Salz11a}, although some of them only exist under conditions too
extreme to be commonly observed terrestrially~\cite{Fuen15a} and some as well
are metastable. Additionally, as thermodynamic conditions change, these
different polymorphs can undergo solid-state transformations from one form to
another.  The common polymorph usually encountered in everyday life called is
\emph{hexagonal ice} (\iceh). Another form, \emph{cubic ice} (\icec)\footnote{
           Emphasizing the uncertainty of the state of knowledge of ice, it has recently
           been suggested that \icec\ is not, in fact, a form of ice that has been observed in
           its pure form. Instead, Malkin~\etal~\cite{Malk15a} contend that previous reports
           of \icec\ are really the disordered form. Whether or not this will be confirmed by
           additional studies, \icec\ gives a convenient boundary condition on the possible
           structures that could exist. We will proceed as though \icec\ does exist, but this
           doesn't affect our discussion or conclusions.},         
has been
observed to coexist with \iceh\ at temperatures as high as
240~K~\cite{Thur13a}.  Above 170~K, \icec transforms irreversibly to \iceh.
There are also structures that have significant disorder, sometimes
called \emph{stacking-disordered ice} abbreviated (\icesd) by Malkins \etal~\cite{Malk12a} and
(\icech) by Hansen \etal~\cite{Hans14a}, with the subscript `ch' in the latter designation indicating that
it is a mixture of \icec\ and \iceh.

\begin{figure*}
\begin{center}
\resizebox{!}{2.4in}{\includegraphics{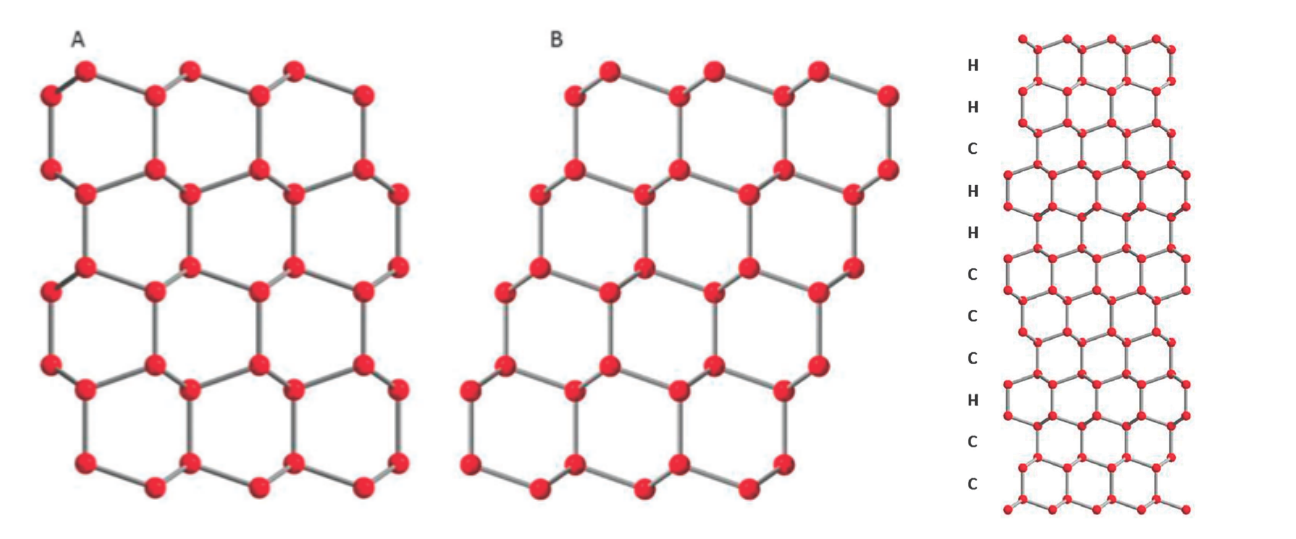}}
\end{center}
\caption{(\emph{left}) The stacking of layers in hexagonal ice (\iceh). The vertical axis is normal to the 
  (0001) basal surface of hexagonal ice. Only oxygen atoms (red spheres), which are connected by 
  hydrogen bonds (gray lines), are shown.  
               (\emph{center}) The stacking of layers in cubic ice (\icec), with the vertical axis normal the (111) plane.               
               (\emph{right}) An example of a stacking sequence that may come from stacking disordered ice (\icesd).
               The layers are marked depending on whether the layer is hexagonally (H) or
               cubically (C) related to its neighbors.
                              Adapted from Malkin \etal ~\cite{Malk15a}. Used with permission.}
\label{Fig:Ice2H3C}
\end{figure*}

Structurally, ice I (\iceh, \icec, \icesd) can be thought of as a
layered material. The oxygens in the water molecules organize into layers
consisting of six-member puckered rings~\cite{Malk12a}\footnote{Note that the
position of the oxygens does not uniquely fix the positions of the hydrogens.
In ice I, of the four possible positions that may be occupied by a hydrogen,
only two are, and these are usually taken to be random. Thus, ice I is referred
to as \emph{proton disordered}. We do not consider proton disordering in our
analysis.}. These layers can further assume only three possible stacking
orientations, called $A$, $B$, or $C$, just as in close-packed
structures~\cite{Orti13a}. The layers are organized so that upon scanning the
material, the layers form \emph{double layers}, where each individual layer in
this double layer must have the same orientation. Additionally, just as in the
close-packed case, adjacent double layers may not have the same orientation.
Since stacking faults are confined to interruptions \emph{between} the double
layers, one usually takes a double layer as a \emph{modular layer}
(ML)~\cite{Silv12a}, and labels it by $A$, $B$, or $C$. Thus, \iceh\ is
given by $\dots ABAB \dots$ (or equivalently $\dots BCBC \dots$ or $\dots CACA
\dots$), and \icec\ by $\dots ABCABC \dots$ (or equivalently $\dots CBACBA
\dots$). It is sometimes more convenient to work with an alternative notation,
called the Wyckoff-Jagodzinski notation~\cite{Orti13a}. One considers triplets
of MLs, and labels the center ML as either $h$ or $c$, depending on whether it
is hexagonally ($h$) or cubically ($c$) related to its neighbors. For example,
the inner most four MLs of the stacking sequence $ABCBCA$ would be written as
$chhc$. It should be apparent that any stacking structure, whether ordered or
disordered, can be expressed as some $hc$-sequence. The \iceh\ stacking
structure is displayed in Figure (\ref{Fig:Ice2H3C}, \emph{left}) and \icec\
is in Figure (\ref{Fig:Ice2H3C}, \emph{center}). A possible disordered stacking
sequence is shown in Figure (\ref{Fig:Ice2H3C}, \emph{right}).

However, despite a recent flurry of theoretical, simulation, and experimental
studies~\cite{Atta12a,Malk15a,Thur13a,Malk12a,Hans14a,Silv12a,Murr15a,Moor11a,Moor11b,Kuhs12a}, 
there is still much that is not understood about the formation of ice
or the transformations between the various polymorphs~\cite{Salz11a}. In an effort to
understand the coexistence of \iceh\ and \icec\ at low temperatures,
Th{\"{u}}rmer and Nie~\cite{Thur13a} examined their formation on Pt via
scanning tunneling microscopy and atomic force microscopy. They find a complex
interplay between initial formation of \iceh\ clusters that grow by layer
nucleation and eventually coalesce. The details of the coalescence and the
nature of domain boundaries between nucleation centers strongly influence whether
subsequent growth is \iceh\ or \icec. Importantly, they demonstrate that
ice films of arbitrary thickness can be imaged at molecular layer resolution.
Several groups~\cite{Malk12a,Hans14a,Malk15a,Murr15a} have applied the
\emph{disorder model} of Jagodzinski~\cite{Jago49a,Jago49b} to simulated or
experimental X-ray diffraction patterns, using a range of influence between
layers, called the \emph{Reichweite}, of $s = 2, 3, 4$. They find that it is
necessary to use $s=4$ to describe some samples. Molecular dynamics
simulations~\cite{Moor11b} showed that ice crystallizing at 180 K contains both
\icec\ and \iceh\ in a ratio of 2:1. While other molecular dynamics
simulation studies~\cite{Pirz11a} found that pairs of point defects can play an
important role in shifting layers in ice I. Yet other molecular
simulations~\cite{Russ14a} suggested that a yet new phase of ice, called
{\emph{ice 0}}, may provide a thermodynamic explanation for some features of
ice growth.

Chaotic crystallography yields important insights into the kinds of appropriate
models and the nature of stacking processes observed, as well as aids in
comparing experimental, simulation, and theoretical studies. In this way,
chaotic crystallography provides a common platform to relate these diverse
observations and calculations.

\begin{figure*}
\begin{center}
\resizebox{!}{2.7in}{\includegraphics{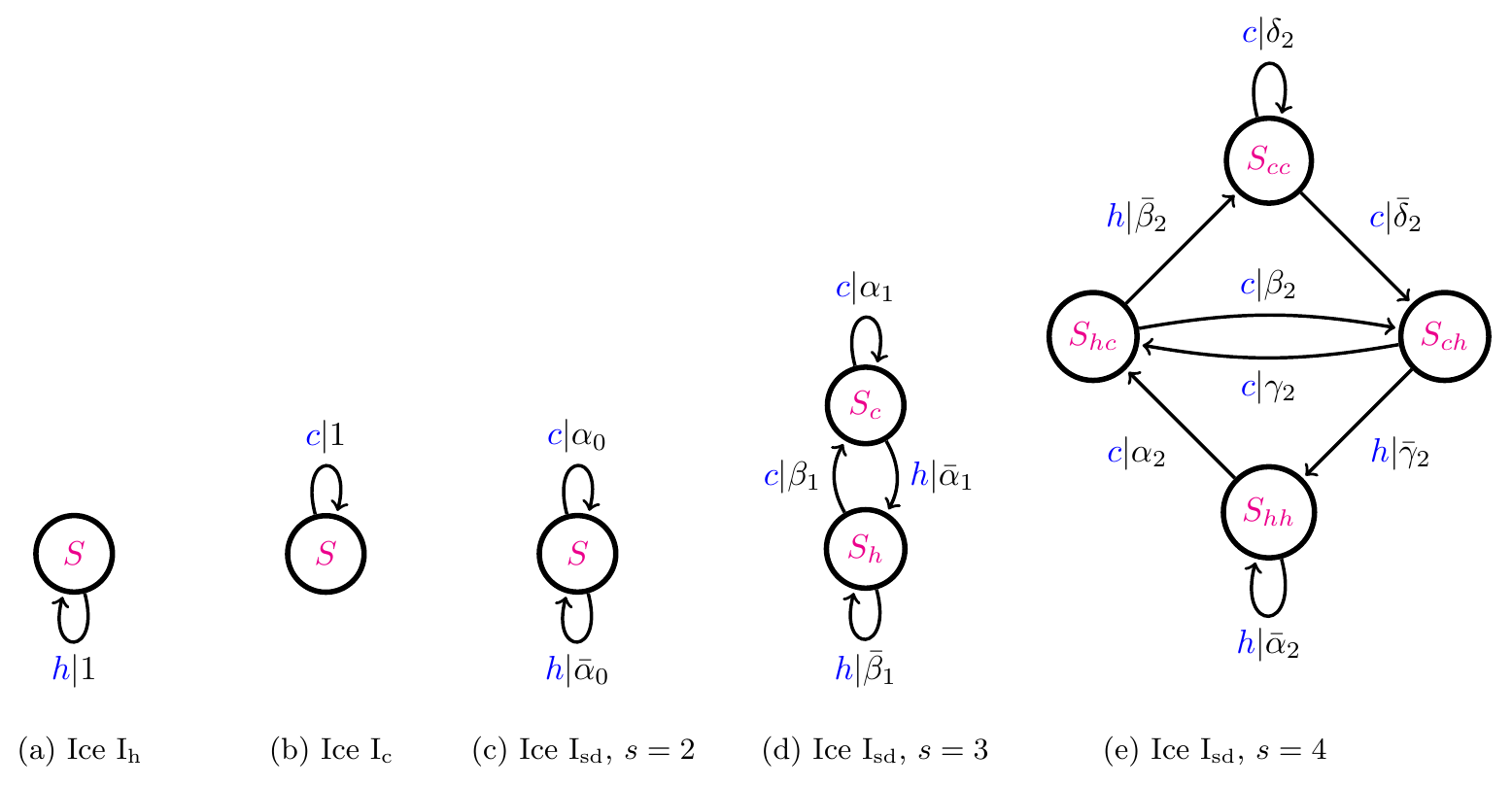}}
\end{center}
\caption{\EMs\ describing the stacking of ice I. Nodes represent causal states and
  are connected by arcs labeled $s|p$, where $s$ is the symbol emitted and $p$
  is the probability of making such a transition. 
      (a) The \eM\ for \iceh\ and               
      (b) \icec.   
      	Models for disordered stacking sequences of close-packed structures
		were introduced by Jagodzinski~\cite{Jago49a,Jago49b}. The model
		parameter specifying the range of influence between MLs is called the
		\emph{Reichweite} $s$.
      (c) The simplest \eM\ in the \emph{hc}-notation that gives an \icesd\
	  	stacking sequence. 
      (d) The \eM\ for $s=3$ \icesd\ and (e) $s=4$ \icesd.}
\label{Fig:IceEmachines}
\end{figure*}

Let's begin with the models used. The \eMs\ that describe \iceh\ and \icec\
are shown in Figs.~(\ref{Fig:IceEmachines}a) and
(\ref{Fig:IceEmachines}b). They are quite similar, both having but one state
and one transition each. Computationally, they are quite simple. Also simple is
the \eM\ shown in Fig.~(\ref{Fig:IceEmachines}c). There are two transitions
from a single state, with the probability of a $c$ being $\alpha_0$ and an $h$
being $\bar{\alpha}_0$.\footnote{Here and elsewhere we adopt the convention
that a bar over a variable means one minus that variable, \ie, $\bar{x} = 1-
x$.} It is apparent that the previous two models are just special cases of
this latter one. We recognize that these three models describe
\emph{independent and identically distributed} (IID) stacking processes. They
imply no correlations between the symbols. However, the coding scheme used
here, the transformation of the $ABC$-notation to the Wyckoff-Jagodzinski
notation, builds in stacking constraints and effectively gives a two-ML
influence distance. We identify this range of influence as the
\emph{Reichweite} $s$.

The next model commonly used is Jagodzinski's $s=3$ disorder model in
Fig.~(\ref{Fig:IceEmachines}d). Here, the next symbol in the sequence depends
only on the previous symbol (either $h$ or $c$), making this a first-order
Markov model. The last model explored in the literature is Jagodzinski's $s=4$
disorder model, and this is depicted in Fig.~(\ref{Fig:IceEmachines}e). Now,
the probability of observing the next symbol depends on the previous two
symbols, we recognize this as a second-order Markov model. Again, the mapping
of the $ABC$-notation to the Wyckoff-Jagodzinski notation folds in an extra
two-ML range of influence in terms of the physical stacking of MLs. It is apparent
that one could continue this process, considering ever larger \emph{Reichweite},
\ie, higher-order Markov models, indefinitely. However, finite-range Markov
processes are only a small fraction of the possible finite-state processes that
one could consider. By finite-state, we mean that there are a finite number of
states; but this does not mean that the range of influence need be finite.
Simulations of simple solid-state transformations in ZnS (also a close-packed
structure) from the hexagonal stacking structure to the cubic one produced
stacking processes with an infinite range of influence~\cite{Varn04a}. Thus, we
are led to suspect that despite the excellent agreement between experimental
and theoretical diffraction patterns reported by some researchers for ice I,
the real process may belong to a computationally more sophisticated class.
Chaotic crystallography, with its emphasis on information- and
computation-theoretic measures, allows one to recognize the possibility and
indeed to ask the relevant questions.

How can we observe or deduce the presence of such sophisticated stacking
processes? One way is improved inference techniques. While chaotic
crystallography has an inference algorithm, \emph{\eM\ spectral reconstruction
theory}~\cite{Varn02a,Varn13a} that detects finite-range processes from
diffraction patterns, there is the possibility of extending it to include
infinite-order processes. Also, the simulation studies discussed earlier can
result in disordered stacking sequences and there are techniques, such as the
subtree merging~\cite{Crut89a} and Bayesian Structure Inference \cite{Stre14a}
algorithms, that can discover these finite-state but infinite-range processes
from sequential data. This suggests that the appropriate level of comparison
between theory, simulation, and experiment is not some signal (the diffraction
pattern), but rather the stacking process itself, as specified by the \eM.
Chaotic crystallography is a platform for such comparison.

Also, by studying the \eM's causal architecture, \ie, the arrangement of causal
states and the transitions connecting them, it is possible to discover the
kinds of faults present. Indeed, this was done for ZnS
polytypes~\cite{Varn02a,Varn07a}. Recently, several different kinds of faults
where proposed for ice I~\cite{Silv12a}, and a proper analysis of the
associated \eM, combined with theoretical and experimental studies, can
elucidate which faults are important in a particular specimen. This could be
quite valuable, as there are many possible routes of formation for disordered
ice specimens, and different mechanisms, such as solid-state transformations
versus growth, likely leave a discernible fingerprint in the causal
architecture.


\subsubsection{Organization of Aromaticity}

Benzene is famous for it's curious ``aromatic'' character that stems directly
from the six $\pi$ electrons shared between its six carbon atoms and hovering
above and below the plane of its carbon-atom ring. To understand this
character, chemists are trying to localize the delocalized $\pi$ electrons,
partly to understand benzene's physical character and partly to find new ways
to control chemical reactivity and discover new synthetic paths. One goal is to
engineer benzene's novel electronic motif to act as a controllable reaction
catalyst. There is an active research program to modify benzene's aromatic
properties by adding on ``bicyclic'' rings outside the main ring. This led to
the creation of tris(bicyclo[2.1.1]hexeno)benzene (TBHB). TBHB's structure is
critical to understanding how to localize benzene's $\pi$ electrons \cite{Rouh96a}.

\begin{figure}
\begin{center}
\resizebox{!}{1.8in}{\includegraphics{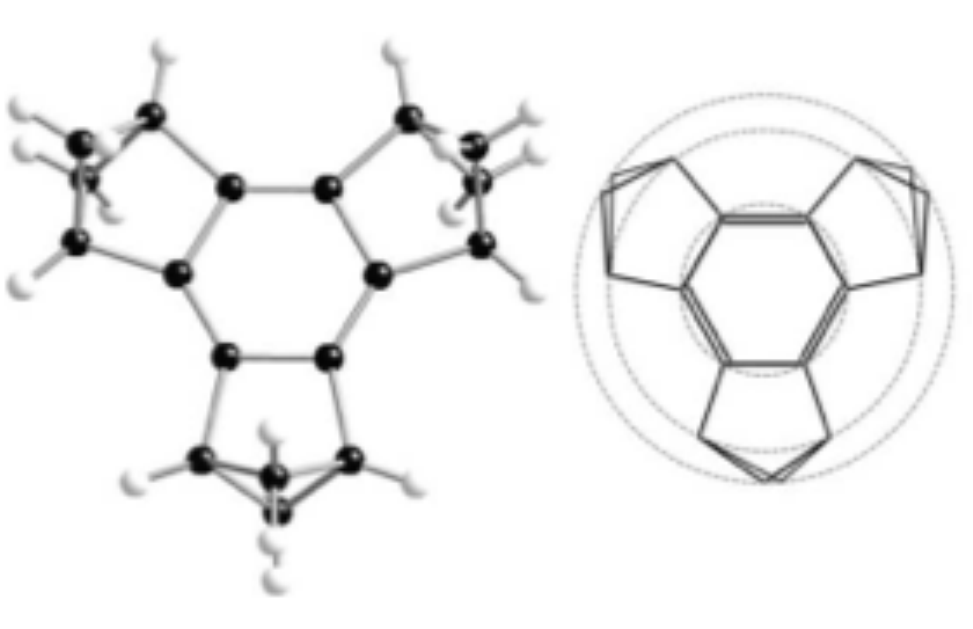}}
\end{center}
\caption{({\it{left}}) Molecular structure of TBHB. Black spheres
  represent carbon atoms while the white are hydrogen atoms. ({\it{right}}) The
  so-called ``skeletal formula'' representation of TBHB. Adapted from
  Michels-Clark \etal \cite{Mich13a}, used with permission.
  }
\label{Fig:MichelsClarkFig1}
\end{figure}

\begin{figure}
\begin{center}
\resizebox{!}{1.8in}{\includegraphics{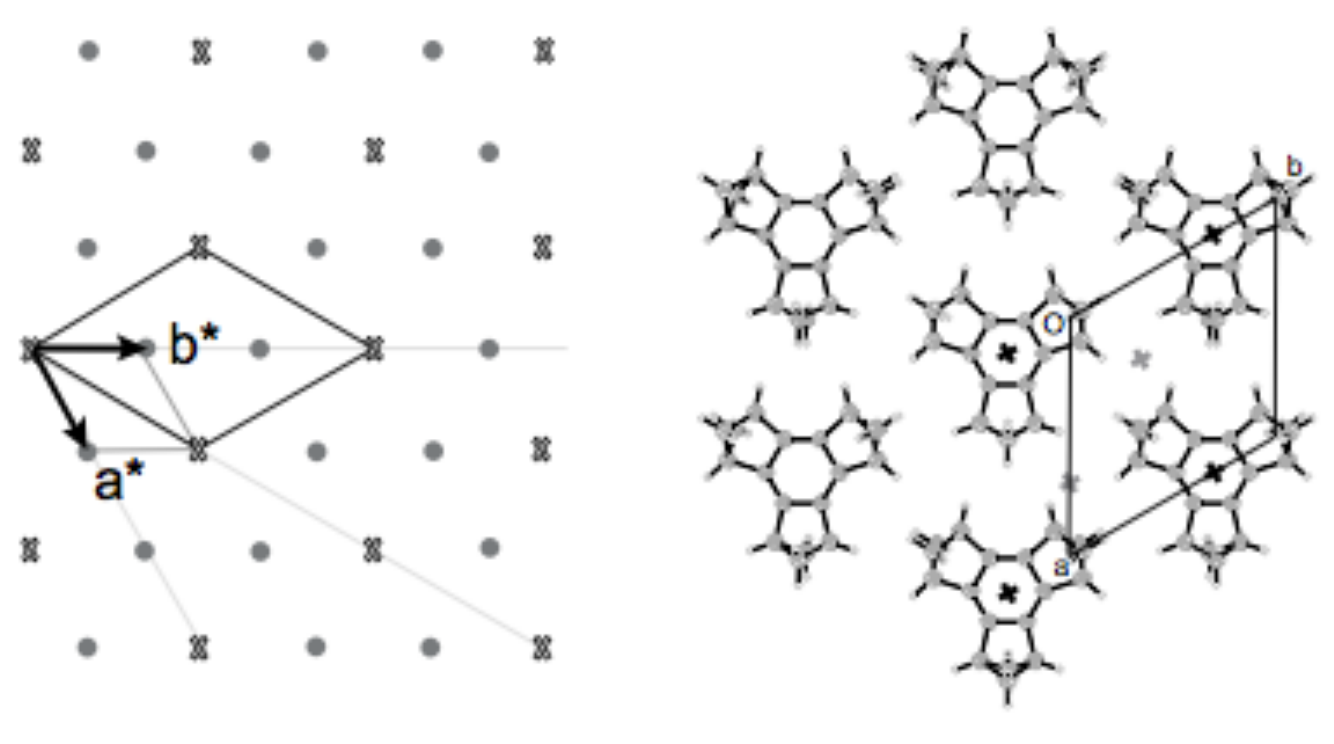}}
\end{center}
\caption{({\it{left}}) Schematic illustration of the reciprocal space plane
  $hk0$. Circles represent the positions of the diffuse scattering ($h-k
  \neq 3n$) and crosses the positions of the Bragg-like reflections
  ($h-k = 3n$).  ({\it{right}}) Average layer structure for TBHB with the
  layer symmetry being $p(\overline{6})2m$. A two-dimensional array of TBHB
  molecules thus arranged is called a \emph{modular layer} (ML). Adapted
  from B{\"{u}}rgi \etal \cite{Burg05a}, used with permission.
  }
\label{Fig:BurgiFig3}
\end{figure}

We recount recent experimental probes of TBHB's structure, demonstrating how an information-theoretic analysis yields additional
insight. TBHB is a largely planar molecule that has attracted attention as one
of the first confirmed mononuclear benzenoid hydrocarbons with a
cyclohexatriene-like geometry~\cite{Burg95a}. Figure~\ref{Fig:MichelsClarkFig1}
({\it{left}}) shows the molecular structure of TBHB, and
Fig.~\ref{Fig:MichelsClarkFig1} ({\it{right}}) gives a schematic formula
representation. Of particular interest is the central benzene ring, where the
internal angles of the carbon-carbon bonds are all 120$^{\circ}$, but there is
remarkable alteration of the two inequivalent bond lengths between the carbons
(1.438(5) - 1.349(6) \AA)\cite{Burg95a}. Of additional interest is the
crystallographic structure of TBHB. Here, two crystal morphologies are observed,
monoclinic and hexagonal\cite{Birk03a}. For this latter structure, X-ray
diffraction studies reveal significant diffuse scattering along rods in
reciprocal space, a hallmark of planar disorder. Figure~\ref{Fig:BurgiFig3}
({\it{left}}) shows the positions of the diffuse rods in reciprocal space, and
Fig.~\ref{Fig:BurgiFig3} ({\it{right}}) gives an illustration of the average
layer structure of TBHB. We will call the extension of this configuration into
a two-dimensional periodic array a ML for TBHB.

Of more recent interest~\cite{Burg05a,Mich13a}, and the problem that concerns us here, is quantifying and describing the disordered stacking structures observed in TBHB. In order to do this, we must specify the possible ML-ML stacking arrangements and establish a convenient nomenclature to express extended stacking structures. The stacking rules and conventions for layers of TBHB can be summarized as follows:\footnote{The stacking rules given here may
	seem nonintuitive. Both Burgi \etal \cite{Burg05a} and Michels-Clark \etal
	\cite{Mich13a} give excellent and extended discussions of the stacking
	possibilities and the geometrical and chemical constraints that cause them.
	We only synopsize those results, and the interested reader is urged to
	consult these references for a detailed explanation.}
(i) While there are three ways that two MLs can be stacked, they are
geometrically equivalent and are related by a rotation of 120$^{\circ}$ about
the stacking direction. Thus, there is only single kind of ML-ML relationship.
(ii) For triplets of MLs, there are two geometrically inequivalent stacking
arrangements. For the case where a molecule in the $(i+2)^{\mathrm{th}}$ ML is
directly above one in the $i^{\mathrm{th}}$ ML, this arrangement is called
\emph{eclipsed}. The other distinct possibility is that the
$(i+2)^{\mathrm{th}}$ ML occupies one of the other two positions. These are
geometrically equivalent, being related by a mirror operation, and are called
\emph{bent}. However, as one advances along the stacking direction, these
latter two can be differentiated as rotating in either a clockwise or an
anticlockwise fashion. Together, then, we need to distinguish between three
different triplets of stacking sequences: an eclipsed triplet, which we
symbolize by $e$, a clockwise bent triplet which we will symbolize by $l$, and
an anticlockwise bent triplet, symbolized by $r$.\footnote{To ease the burden
of the nomenclature we will introduce shortly, we are using \{$l,r$\} for these
triplet sequences, instead of the \{$b_L, b_R$\} used in Michel-Clark \etal
~\cite{Mich13a}. However, there is no change in meaning between these two
sets.} We collect these possibilities into the set $\mathcal{A} = \{r, l, e\}$.

Let us imagine a sliding window that permits observation of but three MLs at a
time. That three-ML sequence is then assigned a symbol from $\mathcal{A}$. The
window then increments along the stacking direction by one ML, so that the last
ML in the sequence becomes hidden, and a new ML is revealed. This new three-ML
sequence can again be specified by one of the symbols in $\mathcal{A}$, such
that the four-ML sequence is given by a two-letter sequence from
$\mathcal{A}$. Thus, a physical stacking sequence can be written as sequence
over the set of these triplets, $\mathcal{A} = \{r, l, e\}$.

Recently, Michels-Clark \etal~\cite{Mich13a} compared three different methods
of determining stacking structure for disordered TBHB from diffraction
patterns: differential evolution, particle swarm optimization, and a genetic
algorithm. Although computationally intensive, they find excellent agreement
between calculated and reference diffraction patterns, obtaining an $R$-factor
fitness of $\overline{R} = 0.0077(3)$ for their best case differential
evolution algorithm. We analyze that case in detail now.

Michels-Clark \etal~\cite{Mich13a} assume a second-order Markov process in the
$rle$-notation\footnote{Note, however, that since each symbol implies
information about the arrangement of three MLs, in terms of the MLs their
stacking process is fourth-order Markovian.}, so that the probabilities of
successive symbols are dependent on only the two previous symbols seen, \ie,
$\mathcal{A}^2 = \{rr, rl, re, lr, ll, le, er, el, ee\}$, which they call
\emph{structural motifs}.  Michels-Clark \etal~\cite{Mich13a} directly report
that the probability seeing $e$ following a pair of $e$'s as
$e_2 = \Pr(e|ee) = 0.008$, which is only two standard deviations above
0. In addition, the probability of the $ee$ sequence itself is only 0.00033.
Thus, we neglect the $ee$ sequence when we construct the hidden Markov model
and take the remaining eight two-symbol histories as our set of causal states:
$\{\mathcal{S}_{rr}, \mathcal{S}_{rl}, \mathcal{S}_{re}, \mathcal{S}_{lr},
\mathcal{S}_{ll}, \mathcal{S}_{le}, \mathcal{S}_{er}, \mathcal{S}_{el}\}$.
Table 1 of Michels-Clark \etal~\cite{Mich13a} relates transition probabilities
between structural motifs to model parameters, so that we can directly
calculate transition parameters from any solution of model parameters. Taking
the values for the best case differential evolution solution given in their
Table 2, we calculate these probabilities. The resulting \eM\ is shown in
Fig.~\ref{Fig:TBHBMachine}. In this way, we can give a chaotic crystallographic
interpretation of that stacking process.

With their relatively large asymptotic state probabilities (0.23) and their
large inter-state transition probabilities (0.97), there is one causal state
cycle~\cite{Varn02a,Varn13a} that dominates the \eM: [$\mathcal{S}_{rr}
\mathcal{S}_{rl} \mathcal{S}_{ll} \mathcal{S}_{lr}$]. This causal state cycle
gives rise to the observed parent structure ($rrll$)$^*$. There are two kinds
of deviations from this structure: (i) those that weakly connect the four
causal states of this dominant causal state cycle, {\emph {intra-causal state
state faults}, and (ii) those that break out of the dominant causal state cycle
and visit peripheral causal states, {\emph {inter-causal state faults.}} 

Of the first kind, one can think of two cases, the insertion of a symbol or the
deletion of a symbol into the normal stacking sequence. The self-state
transition loops on $\mathcal{S}_{ll}$ and $\mathcal{S}_{rr}$ have the effect
of inserting an extra modular layer:
\begin{align*}
  \dots l\,l\,r\,r\,l\,l\,{\underline{l}}\,r\,r\,l\,l\, \dots,
\end{align*}
or 
\begin{align*}
  \dots l\,l\,r\,r\,l\,l\,r\,r\,{\underline{r}}\,l\,l\, \dots,
\end{align*}
where the inserted symbol {\emph {as observed by scanning from the left}} is
underlined. Conversely, the two transitions $\mathcal{S}_{lr} \to
\mathcal{S}_{rl}$ and $\mathcal{S}_{rl} \to \mathcal{S}_{lr}$ have the effect
of deleting a symbol, \ie:
\begin{align*}
  \dots l\,l\,r\,r\,l\,l\,r\,|\,l\,l\, \dots,
\end{align*}
or 
\begin{align*}
  \dots l\,l\,r\,r\,l\,|\,r\,r\,l\,l\, \dots,
\end{align*}
where the pike `$|$' indicates the position of the deleted symbol as the
sequence is scanned from the left. Faulting structures of this kind are often
referred to as \emph{extrinsic} (insertion of a symbol) and \emph{intrinsic}
(deletion of a symbol), respectively. For the specimen analyzed here, there is a
probability of 0.01 that one of these faults will occur in dominant causal
state cycle [$\mathcal{S}_{rr} \mathcal{S}_{rl} \mathcal{S}_{ll}
\mathcal{S}_{lr}$] as each causal state is visited.

\begin{figure*}
\begin{center}
\resizebox{!}{3.0in}{\includegraphics{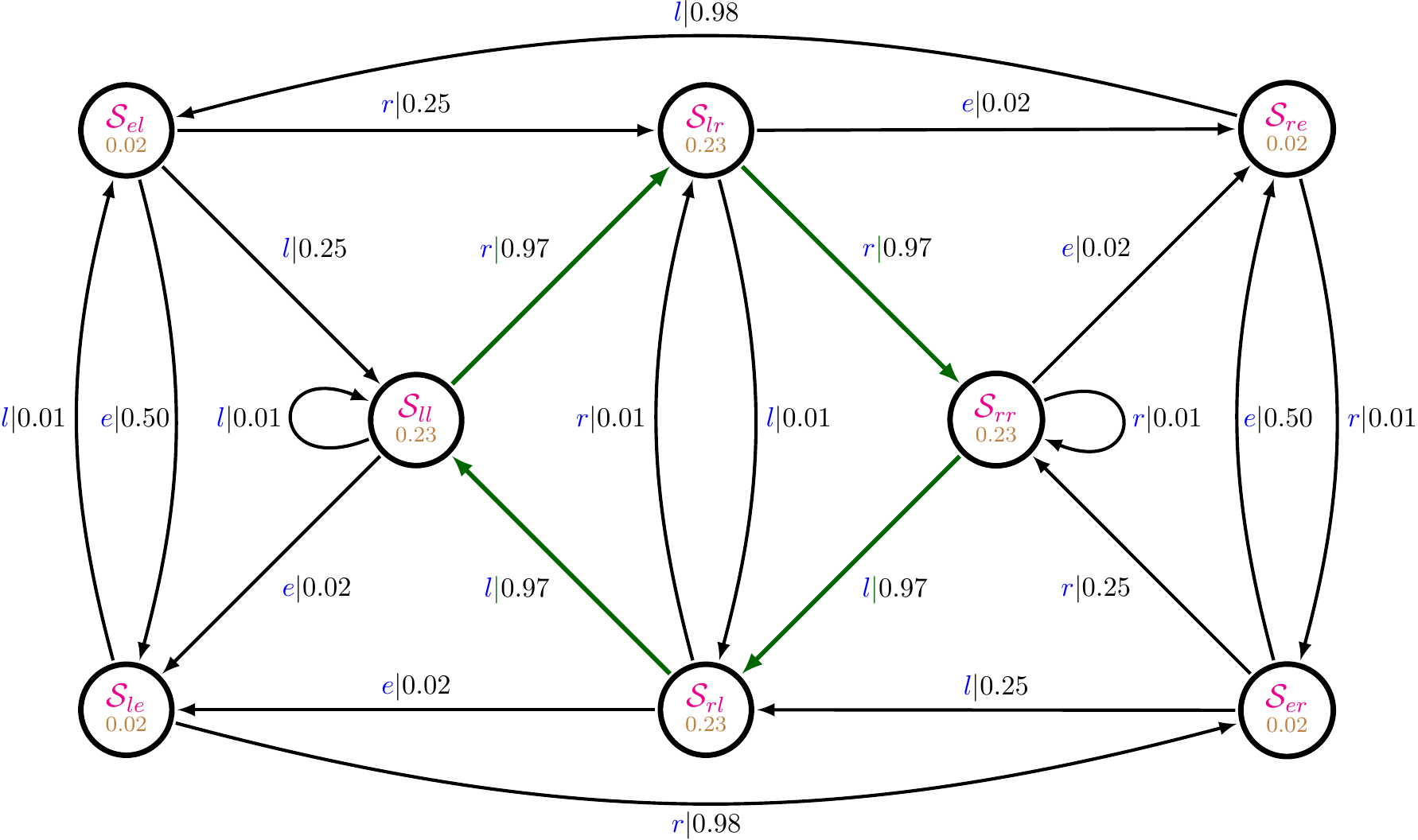}}
\end{center}
\caption{TBHB causal-state architecture. The alphabet is $\mathcal{A}
  = \{l, r, e\}$. In this nomenclature, the process is second-order Markovian
  and, thus, causal states are specified by the last two symbols observed.
  The causal states are labeled with their asymptotic state probabilities.
  There is one closed cycle that dominates the graph, and that is the four-state loop
  [$\mathcal{S}_{rr}\mathcal{S}_{rl} \mathcal{S}_{ll} \mathcal{S}_{lr}$] (green 
  transition arcs). That this nearly periodic sequence is so well represented
  suggests that this structure is nearly crystalline, \ie\ periodic, but does have
  faults.}
\label{Fig:TBHBMachine}
\end{figure*}

A second kind of fault breaks out of this dominant causal state cycle on
emission of an $e$. There is a 0.02 probability of this happening at each
causal state in the dominant causal state cycle. Generically, these transitions
are given by $\mathcal{S}_{yz} \to \mathcal{S}_{ze}$, where $y, z \in \{l,
r\}$. Then, one almost always (0.98 probability) observes that the symbol
following $e$ is {\emph {opposite}} the one that preceded the $e$, \ie:
\begin{align*}
 &\mathcal{S}_{xr} \to  \mathcal{S}_{re} \to  \mathcal{S}_{el}  \,\,\,\, {\mathrm{or}} \\
 &\mathcal{S}_{xl} \to  \mathcal{S}_{le} \to  \mathcal{S}_{er}
  ~.
\end{align*}
In other words, an $e$ is almost always sandwiched between two unlike symbols
drawn from $l$ and $r$. There is a 0.50 probability that the sequence will
return to the dominant causal state cycle at this point on emission of either a
$l$ or $r$. If not, then another $e$ is observed. Thus, even though pairs of
$e$'s ($ee$) are unlikely (and, in fact, prohibited in this \eM) the probability
of observing two $e$'s separated by a single $r$ or $l$ is surprisingly large.
It appears that they may `clump' into small regions.


\subsubsection{Towards a Unified View of Material Structure}
\label{Sec:MatStruct}

How does all of this fit together? Let's contrast the chore of the
crystallographer tasked with determining the structure of a periodic material
and nonperiodic one. For the full three-dimensional periodic case, there are
seven possible crystal systems: \emph{triclinic}, \emph{monoclinic},
\emph{orthogonal}, \emph{tetragonal}, \emph{cubic}, \emph{trigonal}, and
\emph{hexagonal}. One, of course, can be more specific and note that there are
230 crystallographic space groups. A periodic crystal must belong to one and
only one of them. Thus, crystallography is equipped with tools that partition
the space of all possible crystal structures into a finite number of
nonoverlapping sets. Of all the bewildering number ways one might imagine
putting atoms together in a periodic three-dimensional array, this limited
classification system exhausts the possibilities. One can discuss the
similarities between the different systems~\cite{Grim15a} and otherwise
approach a genuine understanding of varieties of possible structures. But can
the same be said for nonperiodic materials?

To simplify the discussion, let's confine our attention to the one-dimensional
case of stacking 1000 MLs. Let's suppose that this is over an alphabet of
two. How many possible stacking sequences are there? Well, there are $2^{1000}
\approx 10^{301}$. Given that there are about $10^{80}$ protons in the
observable universe, it is clear that a comprehensive listing is simply not
possible. And if it were, it is questionable how helpful it would be. For these
disordered materials, then, we are forced to appeal to statistical methods.
Instead of a classification scheme fine-grained at the level of individual
sequences, we instead collect all sequences that have the same statistical
properties into a set. Colloquially, each set is represents a \emph{stacking
process}. Operationally, we attempt to identify to which process a particular
sequence belongs, and then we analyze the process in lieu of the particular
sequence.

Each of the graphs in Fig. \ref{Fig:IceEmachines} and Fig.
\ref{Fig:TBHBMachine} specifies a particular process and defines a hidden
Markov model. While there are still an infinite number of possible processes in
the limit of indefinitely long sequences, a kind of order has been imposed. We
can, for example, enumerate all the processes over a two-symbol alphabet with
just one state. There is but one, and it is shown in Fig.
\ref{Fig:IceEmachines}c. (Figs. \ref{Fig:IceEmachines}a,b are just special
cases of Fig. \ref{Fig:IceEmachines}c.) For two-state binary processes, there
are thirteen~\cite{Stre14a}.\footnote{We brush aside some technicalities here.
In this enumeration, we require that each state transition to only one
successor state on the emission of a particular symbol, a property called
\emph{unifilarity}.  These details, while important, do not detract from our
main point here that process space over a finite alphabet can be systematically
ordered.} For binary processes, the number of distinct processes up to six
states has been tabulated~\cite{John12a}.  Thus, chaotic crystallography does
for disordered materials much the same service that classical crystallography
does for perfectly ordered ones: it organizes and structures the space of
possible atomic arrangements. Further, it allows comparison of the hidden
Markov models between different materials in much the same way that crystal
structures of different materials are compared according to which, for example,
crystal system they belong. 

We contend then that the hidden Markov models describing not only different
specimens of the same material, but different materials altogether can be
compared, either by direct examination of the graphical model of the process or
by information measures that characterize various computational requirements.
As an example, we can compare measures of intrinsic computation between the two
materials considered in the previous subsections as well as that of a third
layered material, ZnS. Of the many measures one can select, we choose
to examine these materials' informational organization via a
\emph{complexity-entropy diagram} \cite{Feld08a}. A complexity-entropy diagram plots,
for each stacking process, the entropy rate $\hmu$ of a symbol sequence
discussed in Section~\S~\ref{InformationMeasures} and the mutual information
between two halves of the specimen, the excess entropy $\EE$, introduced in
Sec.~\S~\ref{sec:ChC}. These measures can be calculated directly from the
hidden Markov model for the stacking processes.

We begin with ice. Note that \icec\ and \iceh\ are both described by single-state
machines and, thus, each half of the crystal shares no information with the
other half, giving $\EE({\rm{I}_c}) = \EE({\rm{I}_h}) = 0$ bits.  Similarly,
being perfectly ordered, we find  $\hmu({\rm{I}_c}) = \hmu({\rm{I}_h}) = 0$
bits/ML. For \icesd, we calculate this quantity for a number of experimental
specimens reported in the literature. Malkin \etal~\cite{Malk15a} 
performed X-ray diffraction studies of several samples of ice I that has been recrystallized from ice II 
and heated at rates between 0.1 to 30 K per minute over temperature ranges of 148 - 168 K.
They use the $s=4$ Jagodzinski disorder
model to analyze their results, and we find by direct calculation from the data given in their Table 4
that these information measures cluster in the range of $\EE({\rm{I}_{sd}}) \approx 0.10 - 0.15$ bits and
$\hmu({\rm{I}_{sd}}) \approx 0.75 - 0.90$ bits/ML. 
Murray \etal~\cite{Murr15a} carried out similar studies on 
ice I deposited as amorphous ice from the vapor phase onto a glass substrate at 110 K.  
The sample was subsequently warmed at a rate of
1 K per minute, and they report diffraction patterns recorded at selected temperatures in the range of 125 - 160 K. 
They too analyzed the diffraction patterns using the $s=4$ Jagodzinski disorder
model, although they found that memory effects were negligible. 
We find by direct calculation from the data given in their Table 1 that these information measures
cluster near $\EE({\rm{I}_{sd}}) \approx 0$ bits and
$\hmu({\rm{I}_{sd}}) \approx 0.95 - 1.00$ bits/ML. 
We can do the same for TBHB.
We find $\EE({\rm{TBHB}}) = 1.9$ bits and $\hmu({\rm{TBHB}}) = 0.27$ bits/ML.
For comparison, we also consider these quantities for several specimens of ZnS
analyzed elsewhere~\cite{Varn07a}. Lastly, to contrast with these disordered samples,
we consider a one-dimensional process that has characteristics similar to
those of a quasicrystal, the \emph{Thue-Morse Process} (TM)~\cite{Badi97a}.
Like a quasicrystal, it is completely `ordered', but nonperiodic.  We have $
\lim_{N \to \infty} \EE({\rm{TM}}) = \infty$ bits, where $N$ is the number of
layers in the specimen, and $\hmu({\rm{TM}}) = 0$ bits/ML. 

Since the maximum possible stacking disorder is 1 bit/ML for ice I, we can see
that disordered ice I really is, well, disordered. Additionally, very little
information ($\EE$) is shared between the different halves. There is little one
can predict about one half of the specimen knowing the other half. The
clustering of the these information measures does lend credibility to the
notion that \icesd\ is a `new' form of ice. We
would, however, exercise caution in referring to this as a distinct
thermodynamic phase of ice. Observe that it is not only not well defined in
stacking-sequence space, \ie\ there are many sequences that correspond to
\icesd, but we also see from the spread of information measures on the
complexity-entropy diagram, it is not well defined in process-space either. We
prefer the interpretation that these specimens are chaotic crystals, each being
described by a different hidden Markov model and each exhibiting different measures of
information processing. Thus, they really do not constitute a separate phase in
the same sense that \icec\ and \iceh\ are. Ice I$_{\rm{sd}}$ is, at least at the moment,
an umbrella term for ice I with a largely randomly stacking of hexagonal and
cubic layers. We do note that information-theoretic measures can distinguish 
between \icesd\ samples having different histories of under different thermodynamic conditions.

In contrast to the near complete disorder of these \icesd\ specimens, the TBHB
sample appears to be much more organized. Not only is more information shared
between the two halves, but the entropy rate is much less. Indeed, as noted
before, there is a prominent cycle on the graph of Fig.~\ref{Fig:TBHBMachine}
that is nearly periodic. ZnS presents an intermediate case. Its
specimens are similar to ice I in that they are either grown disordered or
caught in the transformation between crystalline phases: a hexagonal phase and
a cubic one. Generically, however, ZnS appears to have more structured intermediate states,
suggesting a more structured transformation, likely as a result of significant constraints on the
types disordering mechanisms in play. We can speculate that though ice I and ZnS can both be described as
close-packed structures, the disordering and transformation mechanisms are at
least quantitatively, if not qualitatively, different for each.

\begin{figure}
\begin{center}
\resizebox{!}{3.8in}{\includegraphics{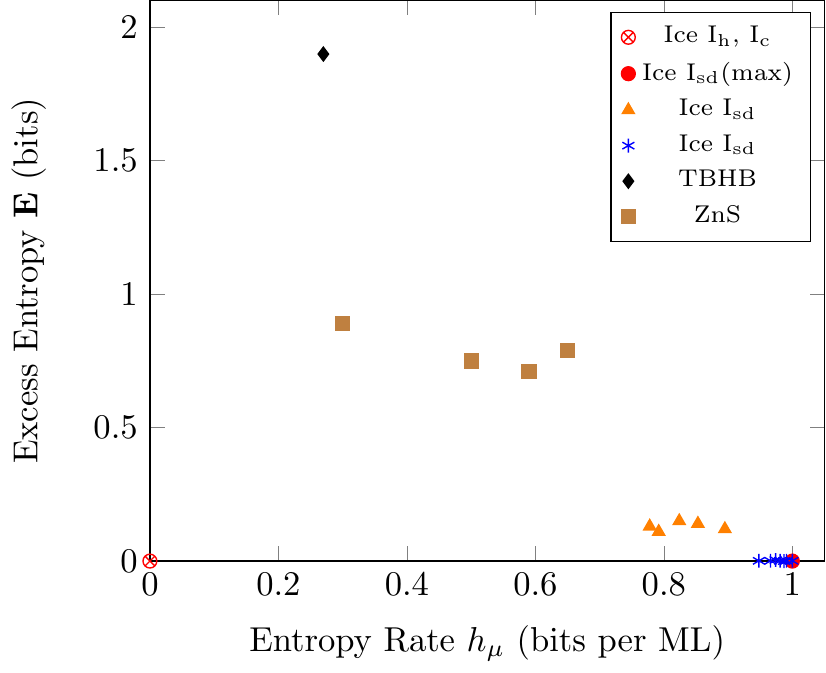}}
\end{center}
\caption{The variety of intrinsic computation as revealed by a complexity-entropy diagram. 
For disordered ice, we plot values calculated from experiments on
I$_{\rm{sd}}$ ({\color[rgb]{0,0,1} $\ast$} from Table 1 of ref~\cite{Murr15a} and
{\color[rgb]{1,0.5,0} $\blacktriangle$} from Table 4 of ref~\cite{Malk15a});
TBHB ($\blacklozenge$ from ref~\cite{Mich13a}); and
ZnS ({\color[rgb]{0.8,0.6,0.4} $\blacksquare$} from ref~\cite{Varn07a}).
For reference, we plot the complexity-entropy point for both 
I$_{\rm{h}}$ and I$_{\rm{c}}$  ({\color[rgb]{1,0,0} $\otimes$}) as well as
I$_{\rm{sd}}$ ({\color[rgb]{1,0,0} \CIRCLE}). Not shown is the 
point for the TM process (see text), a one-dimensional surrogate
for a quasicrystal. We would find it indefinitely high
on the vertical axis (the excess entropy is divergent with a vanishing small entropy rate for large specimens). 
Different values of intrinsic computation indicate significant differences in the organization
of the stacking process for each material as well as their entropy densities.
}
\label{Fig:CompEntr}
\end{figure}

Examining Fig.~\ref{Fig:CompEntr}, we see that the
complexity-entropy diagram also provides a partitioning for the kinds of
structures that can can exist. For example, any periodic process has zero
entropy, thus on a complexity-entropy diagram all perfect crystals are confined
to the vertical axis. This, then, makes concrete just how special crystallinity
is. Similarly, quasicrystals inhabit the upper left corner of the diagram, also
confined to the vertical axis. Thus, while quite interesting, quasicrystals are
informationally rather special organizations. All the space to the right of the
vertical axis is occupied by entropic crystals---just the kinds of specimen
that chaotic crystallography is ideally suited to describe. Thus, chaotic
crystallography introduces tools to quantify these structures and represents a
significant expansion over the domain of classical crystallography.

Although we maintain that understanding structure in itself is a worthy enough
goal, we are mindful that one of the fruits to be harvested from this inquiry is
the possible exploitation of the connection between \emph{structure} and
\emph{function}\footnote{We are reminded here of the 
          well known epigram by Louis Sullivan, 
          ``Form ever follows function''~\cite{Sull96a}. Although uttered in the context 
          of architecture, it applies equally well to materials.}.
The interrelationship between structure and material properties is quite
well known. Carbon can exist as face-centered cubic crystal and, when a
specimen is so ordered, we call it a diamond. More commonly, carbon is found in
hexagonal sheets and is known as graphite. Carbon can also be arranged as
nanotubes and spherical shells informally called \emph{Bucky balls}. And,
though each of these is equivalent by composition, their material properties
are vastly different. Structure matters. Less drastically, different
kinds of stacking structures change material properties in more subtle ways.
Brafman and Steinberger \cite{Braf66a} noticed that by changing from one kind
of periodic stacking structure in ZnS to another, the degree of birefringence
changes. Indeed, this change appeared to depend on only a single parameter, the
hexagonality, which is the fraction of layers hexagonally related to
their neighbors, given by $\Pr(h)$. And, perhaps consequentially, it did so in a very smooth and
predictable way. We know that stacking structure affects other material
properties, such as the diffraction pattern and, clearly, the correlation
functions. It requires little imagination to speculate that other properties
may be similarly affected.

Let us then return to the case of stacking $1000$ MLs. Suppose we
task a materials scientist to investigate the possible material properties
obtainable from different stacking sequences. Even in the simple binary case,
as we saw above, there are approximately $10^{301}$ such sequences. Thus, (the
admittedly naive approach of) a detailed sequence-by-sequence analysis is
unfeasible---either experimentally, theoretically, or via simulation. Yet
absent any theory of disorder in materials, such a brute force investigative
approach might be thought necessary. A chaotic crystallography perspective
immediately equips the materials scientist with tools to approach the problem.
She knows, for instance, that many materials properties are dictated by and
calculable from knowledge of the stacking process alone. Thus, instead of
trying to tackle the problem sequence-by-sequence, it is profitable instead to
approach it process-by-process. Although the space is still enormous, it is
considerably smaller and, importantly, is now systematized. Starting with
simple processes and proceeding to more complex ones, might, for example, be an
effective strategy.\footnote{Chaotic crystallography has tools for quantifying
the complexity of the stacking process\cite{Varn13a,Varn15a}, so this notion of
treating simple processes first can be made operational.} Furthermore,
properties may not even depend on the stacking process details, but may instead
correlate with overall statistical properties or information theoretic
measures. The case of the birefringence of ZnS hints at this. A single
statistical parameter correlates with the observed birefringence; at least for
periodic stacking sequences. Also, the diffraction pattern is known to depend
only on the pairwise correlations between MLs. It is well known that different
stacking processes can have the same correlation functions, suggesting that an
even less fine-grained approach may be profitable. To the extent that
transmission properties through disordered potentials depend only on
correlation functions~\cite{Izra12a}, here too a less fine-grained approach may
be useful.

One may object and question whether we are guaranteed that all material
properties are the same for all realizations of a process. We are not.
However, theoretical results suggesting the important parameters to consider,
coupled with experimental observations and the outcomes of simulations, can
give confidence that a particular property under study is an ensemble
property.  Unquestionably, much of the connection between information-theoretic
properties and materials properties remains unexplored. Along the lines
presented here and paralleling Schr\"odinger's principled, but speculative
thoughts about life's organization, the abundant hints of intimate connections
are too promising and possible rewards of finding and exploiting such
connections are too rich to not explore.

We note too that the exercise of predicting material properties from structure
is by no means academic: The Materials Genome Project \cite{MGI11a} is a
coordinated and dedicated effort spanning theoretical, experimental, and
simulation studies attempting to do just this. Given the sheer variety of
possible arrangements of atoms, an organizational scheme that structures the
space of possibilities is an absolute necessity. Otherwise, researchers will
find themselves relying on intuition---formidable certainly, but all too often
unreliable---alone to propose and assemble possible configurations with novel
material properties. Without too much exaggeration, it is a akin to banging on
a keyboard hoping to finger out one of Shakespeare's sonnets: Possible yes, but
ever so much more likely if one knows the rules of English grammar.

\section{Thermodynamics of Material Computation}
\label{Sec:ThermoComp}

Up to this point, we focused exclusively on informational properties embedded
in the static structure of ``chaotic'' materials, ignoring
temporal dynamics ... of their growth, their functional behavior in the
``wild'', and the like. A full story, though, requires a thermodynamic
accounting of the informational aspects of such materials---the energetics of
their equilibrium and nonequilibrium configurations, the energetics of how they
come to be, how they are transformed, and what functions they support.  Here,
to illustrate the connections between intrinsic information and energetic
costs, we briefly review recent explorations of Maxwell's Demon and a ratchet
model that describes how molecular ``engines'' can store and process
information as they traverse a control sequence.

\subsection{Szilard's Single-Molecule Engine}
\label{Sec:Szilard}

Biological macromolecules \cite{Juli97a,Bust05a,Dunn15a} perform tasks that
involve the simultaneous manipulation of energy, information, and matter.
Though we can sometimes identify such functioning---in the current gating of a
membrane ion channel \cite{Patl91a,Daya05a} that supports propagating spike
trains along a neuronal axon or in a motor protein hauling nutrients across a
cell's microtubule highways \cite{Juli97a}---it is not well understood.
Understanding calls on a thermodynamics of nanoscale systems that operate far
out of equilibrium and on a physics of information that quantitatively
identifies organization and function. At root, we must rectify this functioning
with the entropy generation dictated by the Second Law of Thermodynamics. James
Clerk Maxwell introduced the Demon that now bears his name to highlight the
essential paradox. If a Demon can measure the state of a molecular system and
take actions based on that knowledge, the Second Law can be violated: sorting
slow and fast molecules onto separate sides of a partition creates a
temperature gradient that a heat engine can convert to useful work. In this
way, Demon ``intelligence''---or, in our vocabulary, information
processing---can convert thermal fluctuations (disorganized energy) to work
(organized energy).

In 1929 Leo Szilard introduced an ideal Maxwellian Demon for examining the role
of information processing in the Second Law \cite{Szil29a}; a thought
experiment that a decade or so later provided an impetus to Shannon's
communication theory \cite{Lano13a}. Szilard's Engine consists of three
components: a controller (the Demon), a thermodynamic system (a molecule in a
box), and a heat reservoir that keeps both thermalized to a temperature $T$. It
operates by a simple mechanism of a repeating three-step cycle of measurement,
control, and erasure. During measurement, a barrier is inserted midway in the
box, constraining the molecule either to box's left or right half, and the
Demon memory changes to reflect on which side the molecule is. In the
thermodynamic control step, the Demon uses that knowledge to allow the molecule
to push the barrier to the side opposite the molecule, extracting $\int P~dV =
\kB T \ln 2$ work from the thermal reservoir.  In the erasure step, the Demon
resets its finite memory to a default state, so that it can perform measurement
again. The periodic \emph{protocol} cycle of measurement, control, and erasure
repeats endlessly and deterministically. The net result being the extraction of
work from the reservoir balanced by entropy created by changes in the Demon's
memory. The Second Law is respected and the Demon exorcised, since dumping
that entropy to the heat bath requires a work flow that exactly compensates
energy gained during the control step.

Connecting nonlinear dynamics to the thermodynamics of Szilard's Engine, we
recently showed that its measurement-control-erasure barrier-sliding protocol
is equivalent to a discrete-time two-dimensional map from unit square to itself
\cite{Boyd14b}. This explicit construction establishes that Szilard's Engine is
a chaotic system whose component maps are thermodynamic transformations---what
we now call a \emph{piecewise thermodynamical system}. An animation of the
Szilard Engine, recast as this chaotic dynamical system, can be viewed at
\url{http://csc.ucdavis.edu/~cmg/compmech/pubs/dds.htm}.

What does chaos in the Szilard Engine mean? The joint system generates
information---information that the Demon must keep repeatedly measuring to stay
synchronized to the molecule's position. On the one hand, information is
generated by the heat reservoir through state-space expansion during control.
This is the chaotic instability in the Engine when viewed as a dynamical
system. And, on the other, information is stored by the Demon (temporarily) so
that it can extract energy from the reservoir by allowing the partition to move
in the appropriate direction. To return the Engine to the same initial state,
that stored information must be erased. This dynamically contracts state-space
and so is locally dissipative, giving up energy to the reservoir.

The overall information production rate is given by the Engine's
Kolmogorov-Sinai entropy $\hmu$ \cite{Dorf99a}. This measures the flow of
information from the molecular subsystem into the Demon: information harvested
from the reservoir and used by the Demon to convert thermal energy into work.
Simply stated, the degree of chaos determines the rate of energy extraction
from the reservoir. Moreover, in its basic configuration with the barrier
placed in the box's middle and its memory states being of equal size, the
Demon's molecule-position measurements are optimal. It uses all of the
information generated $\hmu$ by the thermodynamic system: All of the generated
information $\hmu$ is bound information $\bmu$; none of the generated information is lost ($\rmu$ vanishes).

Critically, the dynamical Szilard Engine shows that a widely held belief about
the thermodynamic costs of information processing---the so-called Landauer
Principle \cite{Land61a,Penr70a,Benn82,Land89a,Beru12a}: each erased bit costs
$\kB T \ln 2$ of dissipated energy and the act of measurement comes at no
thermodynamic cost---is at best a special case
\cite{Neum66a,Bark06a,Saga12a,Boyd14b}\footnote{Early on, von Neumann
\cite[Lecture 4]{Neum66a} discussed the general costs of information processing
and transmission without falling into the trap of assigning costs only to
information erasure.}. As the partition location varies and the Demon memory
cells change size, both measurement and erasure can dissipate any positive or
negative amount of heat.  Specifically, there are Szilard Engine configurations
that directly violate Landauer's Principle: erasure is thermodynamically free
and measurement is costly---an anti-Landauer Principle. The result is that the
Szilard Engine achieves a lower bound on energy dissipation expressed as the
sum of measurement and erasure thermodynamic costs. In this, the Szilard Engine
captures an optimality in the conversion of information into work that is
analogous to a Carnot Engine's optimal efficiency when converting a difference in thermal energies to work.

\subsection{Information Catalysts}
\label{Sec:ICatalyst}

Szilard's Engine is one of the simplest controlled thermodynamic devices that
lays bare the tension between the Second Law and functionality of an
information-gathering entity or subsystem (the Demon). The net work extracted
exactly balances the thermodynamic (entropic) costs. This was Szilard's main
point, though we see that his Engine was not very functional, merely consistent
with the Second Law. The major contribution was that, long before Shannon's
information theory, Szilard recognized the importance of the Demon's
information acquisition and storage in resolving Maxwell's paradox.

This allows us to move to a more sophisticated device that uses a reservoir of
information (a string of random bits) to extract net positive work from a heat
reservoir. To set the stage for the thermodynamics we are interested in, but
staying in the spirit of complex materials, let's re-imagine the Szilard Engine
implemented as an enzyme macromolecule whose conformational states implement
the measurement-control-erase protocol. Moreover, let this enzyme traverse a
one-dimensional periodic crystal---say, a strand of DNA---reading its
successive base-pairs to obtain individual Measurement, Control, Erase protocol
commands. The preceding thermodynamics and informational analysis thus apply to
such a molecular engine---an actively controlled system that can rectify
fluctuations, being only temporarily, locally inconsistent with the Second Law.

Let's go one step further, though, to imagine a functional enzyme that over a
thermodynamic cycle extracts net positive work from an information reservoir to
store or release energy as it assembles or disassembles a chain of small
molecular components. In this, we replace the one-dimensional ''control''
molecule with a set of random bits that come into local equilibrium with the
enzyme. As they do, the enzyme's dynamic shifts to catalyze assembling the
components. The shift allows the enzyme to selectively use energy from a
reservoir, say an ATP-rich environment whose molecules the machine accesses
when energy is needed (ATP $\to$ ADP) or given up (ADP $\to$ ATP). Figure
\ref{fig:InfoCatalyst} illustrates the new, functional molecular machine.

In this way, the imagined enzyme acts as an \emph{information catalyst} that
facilitates, via what are otherwise thermodynamically unfavorable reactions,
the assembly of the chain of molecular components. In the 1940s, Leon Brillouin
\cite{Bril49a} and Norbert Wiener \cite{Wien48}, early pioneers in the physics
of information, viewed enzymes as just these kinds of catalysts. In particular,
Brillouin proposed a rather similar ``negative catalysis'' as the molecular
substrate that generated negentropy---the ordering principle Schrodinger
identified as necessary to sustain life processes consistent with the Second
Law. Only much later would such ``information molecules'' be championed by the
evolutionary biologists John Maynard Smith and Eors Szathmary \cite{Mayn98a}.

\begin{figure}
\begin{center}
\includegraphics[height=2in]{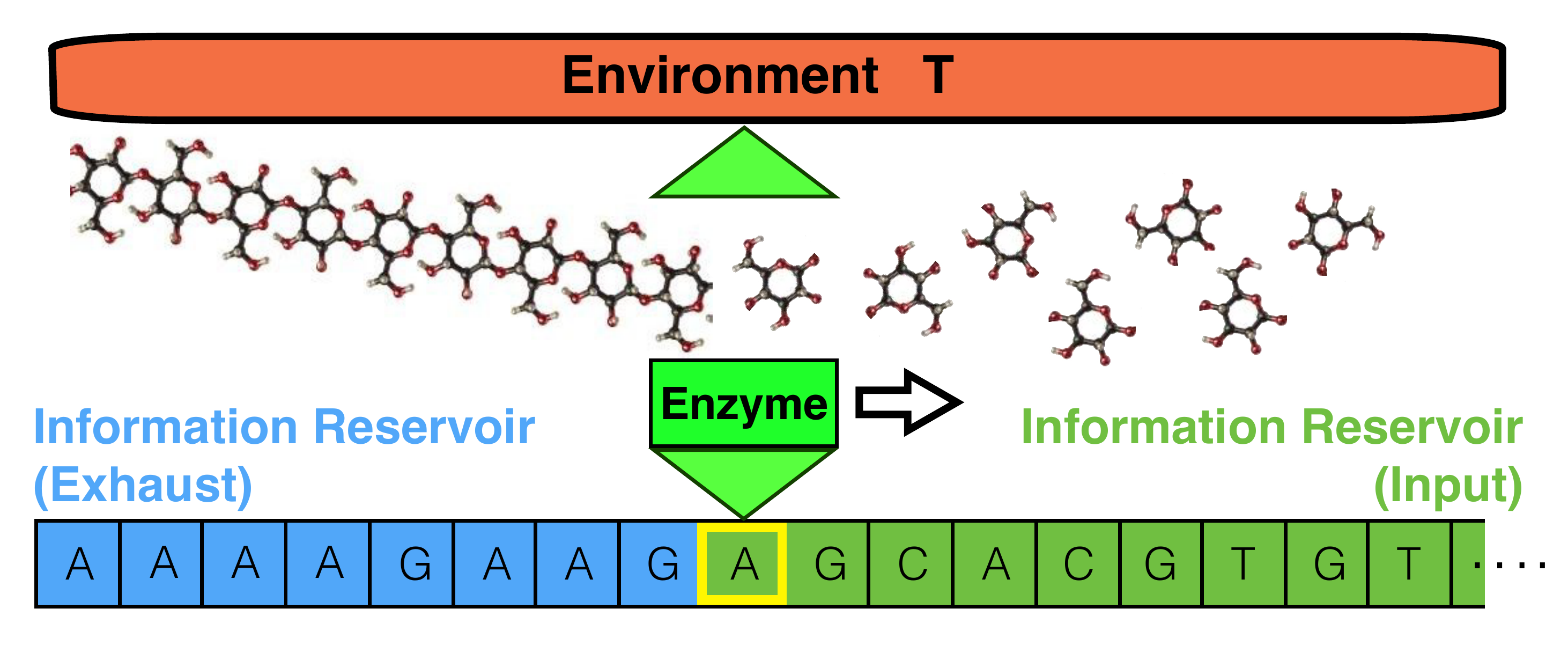}
\end{center}
\caption{Information catalyst: An enzymatic molecular Maxwellian Demon
  thermalizes with a succession of random bits (information reservoir), each
  changing its catalytic activity to overcome energy barriers to assemble a
  chain of simple molecular components. The environment, at constant
  temperature $T$, provides energy-rich molecules needed to drive the
  catalyzed reactions and the chain component molecules.
  (Cf. the information ratchet of Ref. \cite{Boyd15a}.)
  }
\label{fig:InfoCatalyst}
\end{figure}

We recently analyzed the thermodynamics of a class of memoryful information
catalysts \cite{Boyd15a} for which all correlations among system components
could be explicitly accounted. This gave an exact, analytical treatment of the
thermodynamically relevant Shannon information change from the input
information reservoir (bit string with entropy rate $\hmu$) to an exhaust
reservoir (bit string with entropy rate $\hmu^\prime$). The result was a
refined and broadly applicable Second Law that properly accounts for the
intrinsic information processing reflected in the accumulation of temporal
correlations. On the one hand, the result gives an informational upper bound on
the maximum average work $\langle W \rangle$ extracted per cycle:
\begin{align*}
\langle W \rangle & \leq \kB T \ln 2 \, (\hmu' - \hmu)
   ~,
\end{align*}
where $\kB$ is Boltzmann's constant and $T$ is the environment's temperature.
On the other hand, this new Second Law bounds the energy needed to materially
drive transforming the input information to the output information. That
is, it upper bounds the amount $-\langle W \rangle$ of input work to a physical
system to support a given rate of intrinsic computation, interpreted as
producing a more ordered output---a reduction in entropy rate.

This Second Law allows us to identify the Demon's thermodynamic functions.
Depending on model parameters, it acts as an \emph{engine}, extracting energy
from a single reservoir and converting it into work ($\langle W \rangle > 0$)
by randomizing the input information ($\hmu' - \hmu > 0$), or as an
\emph{information eraser}, erasing information ($\hmu' - \hmu < 0$) in the
input at the cost of the external input of work ($\langle W \rangle < 0$).
Moreover, the Demon supports a counterintuitive functionality. In contrast to
previous erasers that only decreased single-bit uncertainty $\H[X_0]$, it
sports a new kind of eraser that removes multiple-bit uncertainties by adding
correlation (temporal order), while single-bit uncertainties are actually
increased ($\H'[X_0] - \H[X_0] > 0$). This modality leads to a provocative
interpretation of life processes: The existence of natural Demons with memory
(internal states) is a sign that they have been adapted to leverage temporally
correlated fluctuations in their environment.

\section{Conclusions}
\label{Sec:Conclusions}

We have come a long way from Schr\"odinger's prescient insight on aperiodic
crystals. We argued, across several rather different scales of space and time
and several distinct application domains, that there is an intimate link
between the physics of life and understanding the informational basis of
biological processes when viewed in terms of life's constituent complex
materials. We noted, along the way, the close connection between new
experimental techniques and novel theoretical foundations---a connection
necessary for advancing our understanding of biological organization and
processes. We argued for the importance of structure and strove to show that we
can now directly and quantitatively talk about organization in disordered
materials, a consequence of breaking away from viewing crystals as only
periodic \cite{Cart12a,Ball14a}. These structured-disordered materials, in
their ability to store and process information, presumably played a role in the
transition from mere molecules to material organizations that became substrates
supporting biology \cite{Cair86a}.  For biology, of course, its noncrystalline
``disorder'' is much much more, it encodes the information necessary for life.
Thus, biological matter is more than wet, squishy ``soft matter''; it is
informational matter. DNA, RNA, and proteins are molecules of information
\cite{Mayn98a,Bril49a,Wien48}. So much so that DNA, for example, can be
programmed~\cite{Adle94a,Adle98a,Winf98a}. And, in a complementary way, the
parallels driving our development here perhaps give an alternative view of
``material genomics'' \cite{MGI11a}.

What distinguishes biological matter from mere physical matter is that the
information in the former encodes organization and that organization takes on
catalytic function through interactions in a structurally diverse environment.
Moreover and critically, these characters are expressed in a way that leads to
increasingly novel, complex chemical structures---structures that form into
entities with differential replication rates \cite{Mayr82a}. And the
high-replication entities, in turn, modify the environment, building ``niches''
that enhance replication; completing a thermodynamic cycle, whose long-term
evolutionary dynamics are thought to be creatively open-ended.

We saw that pondering Schr\"odinger's view of the physical basis of life raised
questions of order, disorder, and structure in one-dimensional materials.
Chaotic crystallography emerged as an overarching theory for the organization
of close-packed materials. It gave a consistent way to describe, at one and the
same time, order and disorder in layer stacking in ice and aromatic compounds
and, generally, in one-dimensional chaotic crystals. And, in this, it hints at
a role that local (dis)ordering can play in enhancing how biomolecules function
synergistically in solution. The issue of biological function forced us to
probe more deeply into its consistency with the Second Law of Thermodynamics.
We then turned to consider two simple cases of Maxwellian molecular Demons to
illustrate that the Second Law of Thermodynamics is perfectly consistent with
the informational character and functionality of smart molecules---that
thermodynamics can begin to describe the energetics of such information catalysts.

Admittedly, we addressed only in a cursory way several major challenges an
informational view of matter poses. Shannon introduced information as surprise
and we showed that this readily led to seeing how information is stored,
created, and transmitted. We only just broached the abiding question, however,
of how these kinds of information contribute to material functionality.
Szilard's Engine and related information catalysts hinted at how we will come
to analyze functional information in complex materials and biomolecules.
Hopefully, the informational perspective will be sufficiently fruitful to
extend to analyzing how such structured objects function in their
environments---how, for example, water plays a critical role in biomolecular
interactions and function.

\vskip6pt

\ack{The authors thank Alec Boyd, Korana Burke, Ryan James, Xincheng Lei,
Dibyendu Mandal, and John Mahoney for helpful conversations. We thank the Santa
Fe Institute for its hospitality during visits. JPC is an SFI External Faculty
member.}

\funding{This material is based upon work supported by, or in part by, the
U~S.~Army Research Laboratory and the U.~S.~Army Research Office under contract
W911NF-13-1-0390.}

\conflict{We have no competing interests.}

\contributions{JPC and DPV wrote the manuscript jointly.}


\bibliographystyle{dpvhack}
\bibliography{wdem}

\end{document}